\documentclass[pre,showpacs,twocolumn]{revtex4}

\newcommand{\gl}[1]{Eq. (\ref{#1})}

\def\gtrless{\raise2.5pt\hbox{$>$}\llap{\lower2.5pt\hbox{$<$}}}
\def\gtrapprox{\raise2.5pt\hbox{$>$}\llap{\lower2.5pt\hbox{$\approx$}}}
\newcommand{\bsq}[1]{\begin{subequations}\label{#1}}
\newcommand{\esq}{\end{subequations}}
\newcommand{\beq}[1]{\begin{equation}\label{#1}}
\newcommand{\eeq}{\end{equation}}
\newcommand{\beqa}[1]{\begin{eqnarray}\label{#1}}
\newcommand{\eeqa}{\end{eqnarray}}

\newcommand{\vek}[1]{{\bf #1}}

\usepackage{amsmath}

\begin{document}

\title{ Light-Scattering by Longitudinal phonons in Supercooled 
Molecular Liquids II: 
Microscopic Derivation of the Phenomenological Equations}
\author{T.~Franosch$^{(1)}$, A. Latz$^{(2)}$, 
R.M. Pick$^{(3)}$}
\affiliation{
$^{(1)}$Hahn-Meitner Institut, D-14109 Berlin, Germany;
$^{(2)}$Technische Universit{\"a}t Chemnitz, D-09107 Chemnitz, Germany;
$^{(3)}$UFR 925, UPMC, Paris, France }
\date{\today} 

\begin{abstract}
The constitutive equations for  the orientational dynamics 
 of a liquid formed of linear molecules 
are derived
microscopically. 
The resulting generalised Langevin equations
coincide with the phenomenological approach of 
Dreyfus {\it et al} \cite{Drey2}. 
Formally exact expressions are given for the phenomenological coefficients 
and various  constraints
 are shown to be consequences 
of this  microscopic approach.


\end{abstract}

\maketitle


\section{Introduction}
Light-scattering has  proven to be an important tool for investigating
condensed matter physics. In the field of supercooled liquids,
 the structural 
relaxation covers many decades either in the time, or in the 
frequency domain, the latter being  
accessible by e.g.,  Fabry-Perot techniques. 
The measured spectra, see e.g. \cite{Goetze1999},  reflect 
the slowing down of the 
structural relaxation upon lowering the temperature and exhibit 
the nontrivial power-laws and stretching effects 
 found by other techniques, such as 
dielectric spectroscopy for instance. 
The most direct measure of the hindered motion
due to the cage effect can be observed by depolarised light-scattering in
the 
back-scattering geometry. For depolarised light-scattering performed at 
other scattering angles,
 one observes an admixture of the transverse 
current motion to the pure back-scattering signal. Similarly,
 for polarised 
scattering,
 one obtains a contribution from density fluctuations which result
in the Brillouin resonance \cite{Dreyfus2002a}.    

The subtlety of light-scattering lies in disentangling the dependence on 
frequency shift, $\omega$, and wave-vector transfer, $q$, as well as on 
incident and outgoing polarisations. 
Since the wave-vector transfer, $q$, is 
small for light-scattering, a generalised hydrodynamics 
approach is suitable. There,
 the spectra are described in terms of a number 
of frequency-dependent memory kernels, e.g. viscosities. 
These kernels have sometimes been written on the basis of heuristic 
arguments. This is the case, for instance of \cite{Drey2}, in which different
previous attempts are also described 
and discussed. A more fundamental approach consists in deriving 
them from a microscopic theory through, say, a Zwanzig-Mori technique. 
The first such attempt was made by Andersen and Pecora \cite{Andersen1971},
who, in fact, made a purely formal use of the technique, the 
memory kernels being eventually approximated by instantaneous 
interactions (Markov approximation). Much more recently, the technique was used
in its full generality in \cite{Franosch2001}. Using only general symmetry
considerations, \cite{Franosch2001} showed that the description 
of the light scattering spectra involved 
 ten frequency-dependent 
functions. This large number  was
the price  to be
paid in order  not to miss any effect that leads to a 
fluctuation of the dielectric tensor $\delta \epsilon_{ij}(\vek{q},t)$.
In the present paper, we shall follow an intermediate route;
we shall derive, for a selected set of dynamical variables, 
 the precise form of  their equations of motion, and of the 
corresponding relaxation kernels. Our results
are   valid whatever the
temperature but  are restricted  to the case of molecular 
supercooled liquids formed of symmetric top molecules. 
Their application to the
light scattering problem requires a precise 
form of $\delta \epsilon_{ij}(\vek{r},t)$; 
following Eq. (12) of Part I of this 
series of papers, we shall {\it assume} $\delta \epsilon_{ij}(\vek{q},t)$ 
to depend  only on two variables of the problem, namely 
 the density and the orientation fluctuations.  
Then the light-scattering 
problem is reduced to calculating the density-density, 
orientation-orientation as well as mixed correlation functions which are 
expressed with the help of appropriate memory kernels. 

As  a further simplification, we shall
ignore temperature fluctuations, i.e. the 
hydrodynamic poles associated with energy conservation. 
This restriction 
is probably justified 
for Brillouin scattering experiments, since, for 
the scattering vectors involved, the  
Rayleigh
line lies at so low a frequency 
that it is  inaccessible to
 the usual  
frequency domain methods. 
Furthermore, for liquids, the ratio of the
isobaric heat capacity to the isochoric one is close to unity and, 
correspondingly, the total weight of the Brillouin lines is much 
larger than the weight contained in the Rayleigh line;  
similarly,  the isothermal sound velocity is close to the adiabatic one.
The situation is different for time-based methods like impulsive thermal 
stimulated Brillouin scattering 
\cite{Yang95,Paolucci2000,Torre2001} 
 where the heat diffusion contribution 
can be observed as a late stage of the relaxation signal, but this aspect of 
the problem will not be dealt with here.

The goal of this paper is twofold. The first is
 to give a microscopic derivation 
of the constitutive equations
 for the density and orientation fluctuations 
used in \cite{Drey1,Drey2,Dreyfus2002a}, and to derive some new
results from this microscopic approach 
\footnote{A very brief study along similar lines was previously reported in 
\cite{Latz2001}.}. The second is 
to compare the results one can obtain from the three
microscopic approaches \cite{Andersen1971,Franosch2001} and the present
paper, which differ in the variables 
taken into account and/or in the scattering model.

Consequently, this paper is organized as follows. 
The phenomenological equations of \cite{Drey1,Drey2} are microscopically 
derived in Section II. In particular, we show that the four memory
functions which enter into those equations, namely the bulk viscosity, 
$\eta_b(t)$, the center-of-mass shear viscosity, $\eta_s(t)$, the
rotational friction $\Gamma'(t)$, and the rotation-translation 
coupling, $\mu(t)$, can be expressed in terms of the dynamical variables 
of the problem, and of a reduced time evolution $R'(t)$ which 
does not contain
 the hydrodynamics poles of the problem. Similarly, the rotation-translation
coupling constant, $\Lambda'$, and the molecular libration frequency, 
$\omega_0$, which are the other ingredients of these equations of motion, 
will be expressed in terms of equal time thermal 
averages of some variables of the problem. We make use of these
microscopic expressions of the memory 
functions in Section III to derive necessary conditions on the 
imaginary part of their Laplace transform, and on some 
contributions of them. These conditions will be such that the light scattering 
spectra will be always positive whatever the values of the coefficients
linearly coupling the density and orientation fluctuations to 
$\delta \epsilon_{ij}(\vek{q},t)$. Section IV makes use of the same 
expressions of the memory functions to relate, through a Green-Kubo
formalism, the correlation functions of some variables 
to specific combinations of the Laplace transforms 
of these memory functions. In particular, we
shall show that $\eta_b(t)$ can be expressed as such a correlation function.
The same will be true for $\eta_T(t)$, the memory function which 
takes into account all the retardation effects related to the
propagation of the transverse phonons; this is not a priori obvious
because $\eta_T(\omega)$ will turn out to depend
in a complex way on the
Laplace transforms of several memory functions
defined above, as well as on $\Lambda'$ and $\omega_0$.
 Section V will compare the 
expressions for the light   scattering intensities 
that can be obtained using the three sets of variables and of dielectric
fluctuation models already mentioned. We shall show that
the set proposed in \cite{Andersen1971} leads to awkward forms of the
relaxation kernels when they are not restricted to a Markov 
approximation, but used for a molecular supercooled liquid. 
Conversely, as
expected, the results 
obtained in Part I are a restriction of those of \cite{Franosch2001} 
corresponding to definite simplifying assumptions. A brief 
summary and some comments conclude the paper.

\section{A Zwanzig-Mori derivation of the dynamical equations}
We consider a dense liquid of $N$ linear molecules of mass $m$ 
at temperature $T$ enclosed in a volume $V$. 
Statistical correlations of phase space variables in 
terms of the Kubo scalar 
product \cite{Forster75},
$(A(t) | B) = \langle \delta A(t)^* \delta B \rangle $, 
$\delta A = A - \langle A \rangle$, provide the simplest 
information on the system's dynamics with  
 $\langle . \rangle $ denoting canonical averaging. The thermodynamic  
limit, $N \to \infty$, with fixed particle density, $n= N/V$, 
is implied throughout. 
The time evolution of the observables 
is driven by the Liouvillian ${\cal L}$: $\partial_t A = i {\cal L} A = 
\{ H, A \} $, where $H$ denotes the Hamilton function 
and $\{ \, , \, \} $ the Poisson bracket. 
We consider the dynamics of the fluctuating molecular  
orientation tensor, written directly in the reciprocal space:
\begin{equation}
Q_{ij}(\vek{q}) = N^{-1/2} 
\sum_{ \alpha =1}^N \left( \hat{u}_{\alpha i} \hat{u}_{\alpha j} 
- \frac{1}{3} \delta_{ij} \right) 
e^{ i \vek{q} \cdot \vek{R}_\alpha} 
\, ,
\end{equation}
where the degrees of freedom of the $\alpha$-th molecule are specified 
by  a unit vector, $\hat{\vek{u}}_\alpha $, for the  orientation 
and by the position of its   center-of-mass,  
$ \vek{R}_\alpha$. The spatial modulation of a fluctuation is 
characterized by its wave vector, $\vek{q}$,
 and latin indices denote cartesian 
components. 
The 9 components of $Q_{ij}(\vek{q})$ are not independent, 
since the orientation 
tensor is symmetric and traceless, reducing the number of 
independent components to 5.  
The normalisation is chosen 
such that the correlation functions are intensive.

Furthermore, we consider the fluctuations in the  mass density: 
\begin{equation}
\rho(\vek{q}) = m N^{-1/2} \sum_{ \alpha=1}^N 
\exp \left( i \vek{q} \cdot \vek{R}_\alpha \right) \, ,
\end{equation}
and the cartesian components of the mass current:
\begin{equation}
J_i(\vek{q}) = N^{-1/2} \sum_{\alpha=1}^N  P_{\alpha i} 
\exp \left( i \vek{q} \cdot 
\vek{R}_{\alpha} \right) \, ,
\end{equation}
where $\vek{P}_\alpha$ denotes the momentum of the $\alpha$-th molecule. 
(Equivalently, one could
 use the particle density $n(\vek{q}) = \rho(\vek{q})/m$ 
and the  velocity $v_i(\vek{q}) = J_i(\vek{q})/ \rho_m$, 
where $\rho_m = m n $ is the mean mass density).

\subsection{Static averages}
The static correlation functions need to be evaluated to 
lowest order in $q$ only. Since the Hamilton function respects
 rotational 
invariance, 
all static averages in the liquid phase 
have to remain unchanged under
any rotation of the system: this implies that, e.g. correlators between, 
say, any second rank traceless tensor and any scalar will
 vanish in the long-wavelength limit (see, e.g. \gl{Qrho}).  
 
The static average of  the density can be expressed as: 
\begin{equation}
\label{rhocorr}
(\rho(\vek{q}) | \rho(\vek{q}) ) = m^2 v^2/ c^2 + {\cal O}(q^2)\, ,
\end{equation}
where $c$ is, here, the isothermal sound velocity 
\footnote{See Section IV-B for a careful discussion on the 
isothermal/adiabatic property.}
 defined
 in terms of 
the long-wavelength limit of the 
static structure factor via $c^2 = v^2/S(q\to 0)$, 
while $v = \sqrt{k_B T/m}$ denotes the thermal velocity. 
As usual, the current correlations read: 
\begin{equation}\label{Jcorr}
(J_i(\vek{q}) | J_k(\vek{q}) )= \delta_{ik} m^2 v^2 \, .
\end{equation}

To lowest order in  $q$, 
the equal-time correlators of the tensor variables read:
\begin{subequations}
\begin{equation}
\label{S}
( Q_{ij}(\vek{q}) | Q_{kl}(\vek{q})) = S^2 \Delta_{ij,kl}
+ {\cal O}(q^2) \, ,
\end{equation}
where:
\begin{equation}
\Delta_{ij,kl} =\left( \delta_{ik} \delta_{jl} + \delta_{il} \delta_{jk} 
- \frac{2}{3} \delta_{ij} \delta_{kl} \right) \, 
\end{equation}
\end{subequations}
is a  fourth-rank tensor, the structure of which  is governed 
by rotational symmetry. 
The long-wavelength limit of the 
$9\times 9$ correlators 
in \gl{S} is thus determined by a single number, $S^2$, denoting
 the long-wavelength limit of 
the corresponding generalised structure factor, a quantity which is, as 
in Eq. (\ref{rhocorr}), 
proportional in leading order to $k_B T$.

Due to rotational symmetry,  
the overlap of the tensor variables with the 
density vanishes in the long-wavelength limit according to:  
\begin{equation}
\label{Qrho}
(Q_{ij} (\vek{q}) | \rho(\vek{q}) )= 
 {\cal O}(q^2) \, .
\end{equation}

We shall also need to consider
the tensor currents, $\dot{Q}_{ij}(\vek{q}) = i {\cal L} Q_{ij}(\vek{q})$, 
which 
are normalised by:
\begin{equation}
\label{Omega}
( \dot{Q}_{ij}(\vek{q}) | \dot{Q}_{kl}(\vek{q})) = \Omega^2 \Delta_{ij,kl}
\end{equation}
with the characteristic frequency scale 
$\Omega$; we show, in Appendix A, Eq. (\ref{evOm}), 
that $\Omega = \sqrt{2 k_B T/5 I}$, where $I$ is the moment 
of inertia of the molecule for a rotation around an axis  
perpendicular to the molecule symmetry axis and passing through its center 
of mass. The ratio of the static averages 
of the orientation and the orientational current, 
$\omega_0 = \Omega/S$, will
determine the axial libration frequency, a frequency 
 characteristic of the short-time expansion 
 for the orientation
correlation function (see \gl{omega}). 
Hence, there is a close analogy between the set of Eqs. (\ref{S}) and 
(\ref{Omega}) and the set of 
density plus momentum current correlators whose ratio determines the 
isothermal sound velocity $c$ characteristic of the initial decay 
of the density correlators; $c$ and $\omega_0$ are, to leading order, 
independent of temperature.

The correlation function between the mass current
 and the tensor current
components has now to be considered; it  
is strictly equal to zero, whatever
is $\vek{q}$:
\begin{equation}
\label{cross}
( \dot{Q}_{ij}(\vek{q}) | J_k(\vek{q}) ) = 0 \, .
\end{equation} 
This is 
due to the fact that we put the point of reference of each molecule, 
$\vek{R}_\alpha$, 
at its  center-of-mass, 
(see Appendix A for a thorough discussion of this property).  
The remaining static averages between the four distinguished variables
$\rho(\vek{q}), J_i(\vek{q}), Q_{ij}(\vek{q})$ and $\dot{Q}_{ij}(\vek{q})$ 
vanish due to time reversal symmetry.

\subsection{Constitutive Equations}

The mass conservation law relates the density to the  momentum current: 
\begin{equation}\label{massconservation}
\partial_t \rho(\vek{q},t) = i q_k J_k(\vek{q},t) \, .
\end{equation}
Similarly, the conservation  of momentum  yields: 
\begin{equation}\label{momentumconservation}
 \partial_t J_k(\vek{q},t) =  i q_l \Pi_{kl}(\vek{q},t) \, ,
\end{equation}
where $\Pi_{kl}(\vek{q},t)$ denotes the fluctuating momentum current tensor. 
At last, we can write the trivial identity:
\begin{equation}\label{ddotQ}
\partial_t^2 Q_{ij}(\vek{q},t) = \ddot{Q}_{ij}(\vek{q},t) \, ,
\end{equation}
which defines $\ddot{Q}_{ij}(\vek{q},t)$ as an orientational
 tensor force. 
In order to close the system, we need  constitutive equations for 
the momentum current tensor, $\Pi_{kl}(\vek{q},t)$ 
and the orientational tensor force.
This will be achieved here through  generalised Langevin equations
which will introduce  
appropriate memory kernels. 
Let us first introduce the projection operator, $P$:
\begin{eqnarray}\label{project}
P & = & 
| Q_{kl}(\vek{q}) ) \frac{1}{2 S^2 } ( Q_{kl}(\vek{q}) | \nonumber \\
& & +
| \dot{Q}_{kl}(\vek{q}) ) \frac{1}{2 \Omega^{2} } ( \dot{Q}_{kl}(\vek{q}) | 
+ | \rho(\vek{q}) ) \frac{c^2}{m^2 v^2} (\rho(\vek{q}) | \nonumber \\
& & 
+ |J_k(\vek{q}) ) \frac{1}{m^2 v^2} (J_k(\vek{q}) |
+ {\cal O}(q^2) \,  ,
\end{eqnarray} 
where the 
sum over repeated indices is implied. $P$ is a projection operator
because, once the symmetric character of $Q_{ij}(\vek{q})$ and 
$\dot{Q}_{ij}(\vek{q})$  has been taken into account, 
one can check that indeed $P^2 = P$. $P$
 projects onto the subspace spanned by density, mass current and the 
symmetric traceless parts of the orientation and the corresponding current. 

The time evolution operator, $R(t) = \exp( i {\cal L} t)$, can be exactly
reformulated as 
\begin{equation}
\label{opid}
R(t) = R(t) P  + \int_0^t R(s) P i {\cal L} R'(t-s)  ds + R'(t) \, ,
\end{equation}
with the reduced operator $R'(t) = Q \exp(i Q {\cal L} Q t) Q$, where 
$Q = 1- P$, and a short proof of \gl{opid} is given in Appendix B. 
The benefit of this procedure 
lies in the following. The 
time evolution operator, $R(t)$, 
possesses, in addition to a non-hydrodynamic  part,  
long-lived hydrodynamic modes that are 
due to conservation laws. This leads to resonances 
in the spectra, viz. the Fourier transforms of the 
time correlation functions of all the distinguished variables for small but 
nonzero wave vectors. 
Conversely, the reduced time evolution operator, $R'(t)$, 
devoids the hydrodynamic singularities and correlation 
functions with $R'(t)$ 
are regular in the long-wavelength limit. 
The problem of handling the slow 
relaxation due to hydrodynamic conservation laws is treated explicitly in
the low-dimensional subspace of the distinguished variables. 
On the contrary, 
the slow 
structural relaxation will be dealt with the help of
 correlation functions of $R'(t)$, the second term of the r.h.s. of 
\gl{opid}, which will appear in the form of  memory kernels.  
In the  spirit of generalised hydrodynamics, the long-wavelength properties 
are described properly by keeping the wave-vector dependences introduced 
explicitly by the conservation laws, while the memory kernels can be 
evaluated in 
their long-wavelength limit.    

Before deriving the constitutive equations from \gl{opid} for the missing 
quantities, $\Pi_{ij}(\vek{q},t)$ and 
$\ddot{Q}_{ij}(\vek{q},t)$, some comments on 
the structure of the resulting equations in the long-wavelength limit 
are in order. First, from time reversal symmetry,
 the instantaneous coupling (first term of the r.h.s. of \gl{opid})
will be non-zero only for variables of identical time-parity. Since 
both the momentum current and the orientational force have even time parity,  
this instantaneous part will consist of density and orientation only. 
Second, rotational 
symmetry implies that the coupling of irreducible tensors of different 
ranks
 is suppressed 
in the long-wavelength limit by appropriate powers of the wave number. 
The dynamical correlators enjoy the same property, since the time 
evolution does not change the rank of a  tensor. 
We shall
keep only the lowest nontrivial terms in this small-wave-number expansion, as 
was already hinted at by keeping only 
the lowest order of the static averages in
the preceeding section.

Finally, in the second term of the r.h.s. of \gl{opid},
one can let $i{\cal L}$ operate on the 'bra' part of the projector, $P$;  
for instance:
\begin{subequations}\label{pressuresub}
\begin{eqnarray} 
& &| \rho(\vek{q})) \frac{c^2}{m^2 v^2} (\rho(\vek{q}) | i {\cal L}
  =   - | \rho(\vek{q})) \frac{c^2}{m^2 v^2} (\dot{\rho}(\vek{q}) | 
\nonumber \\
& & =    | \rho(\vek{q})) \frac{c^2}{m^2 v^2} i q_k (J_k(\vek{q}) | \, ;
\end{eqnarray}
because $R'(t-s)$ contains, on its l.h.s, a $Q=1-P$ factor, the contribution
of $J_k(\vek{q})$, and similarly of $\dot{Q}_{kl}(\vek{q})$, are eliminated
from this second term and one obtains:
\begin{eqnarray}\label{PiLR}
& & P i {\cal L} R'(t-s) =   - | \dot{Q}_{kl}(\vek{q}) ) \frac{1}{2 \Omega^2}
 (\ddot{Q}_{kl}(\vek{q}) | R'(t-s) \nonumber \\ 
& & +
| J_k(\vek{q}) ) \frac{i q_l}{m k_B T} (\Pi_{kl}(\vek{q}) | R'(t-s) \, .
\end{eqnarray}

Let's first handle the momentum current
 tensor which we decompose into: 
\begin{equation}
\Pi_{ij}(\vek{q},t) = 
\delta_{ij} p(\vek{q},t) + \pi_{ij}(\vek{q},t) \, .
\end{equation}
Here: 
\begin{equation}
p(\vek{q},t) = [\Pi_{xx}(\vek{q},t) 
+ \Pi_{yy}(\vek{q},t) + \Pi_{zz}(\vek{q},t) ]/3 
\end{equation} 
\end{subequations}
denotes the fluctuating pressure so that  $\pi_{ij}(\vek{q},t)$ 
is a traceless symmetric second rank tensor.
Multiplying \gl{opid} from the right by $p(\vek{q})$ yields the 
desired Langevin equation for 
the pressure fluctuation:
\begin{eqnarray}
\label{Langevin}
p(\vek{q},t) & = & R(t) P p(\vek{q})  
+ \int_0^t R(s) P i {\cal L} R'(t-s) p(\vek{q})  ds \nonumber \\
& & + R'(t) p(\vek{q}) \,  .
\end{eqnarray}
The first term represents an instantaneous coupling to the 
distinguished variables of the projector. The second yields a
retarded coupling and the last term is a rapidly fluctuating 
term that we shall call 'noise', i.e. which  is uncorrelated
for all times to the distinguished variables. Hence, this term can be ignored
for the evaluation of the correlation functions of the distinguished variables.
Nevertheless, the same term
 will be useful in establishing the Kubo formulae 
of Section IV, which will relate the time-dependent correlation 
functions of some variables to the memory kernels of the dynamical equations. 
  
In order to evaluate the first term of \gl{Langevin}, 
we need static correlations of the pressure with the distinguished variables. 
Time-inversion symmetry allows non-vanishing correlations 
only with the density and the orientational tensor. 
Since rotational invariance implies: 
\begin{subequations}
\begin{equation}\label{rotinv} 
(\pi_{ij}(\vek{q}) | \rho(\vek{q})) 
= {\cal O}(q^2) \, ,
\end{equation} 
 one can evaluate $(\rho(\vek{q}),p(\vek{q}))$ by
 using the conservation of momentum, \gl{momentumconservation},
and \gl{Jcorr}, up to terms of order ${\cal O}(q^2)$:  
\begin{eqnarray}
  (\rho(\vek{q}) | p(\vek{q}) ) i q_k  
& = &  
 (\rho(\vek{q}) | p(\vek{q}) ) i q_l \delta_{kl}  \nonumber \\
 &= &  ( \rho(\vek{q})| i  q_l \Pi_{kl}(\vek{q}) ) 
 =  (\rho(\vek{q}) | \dot{J}_k(\vek{q}) )   \nonumber \\
& =& - (\dot{\rho}(\vek{q}) | J_k(\vek{q}) ) = i q_l (J_l(\vek{q}) | 
J_k(\vek{q}) ) \nonumber \\
&= & i q_k m^2 v^2 \, .
\end{eqnarray} 
This yields:
\begin{equation}\label{prhocorr}
( \rho(\vek{q}) | p(\vek{q}) ) = m^2 v^2 + {\cal O}(q^2) \, .
\end{equation}
\end{subequations}

Conversely, rotational invariance implies, similarly to \gl{Qrho}, that 
$(p(\vek{q}) | Q_{ij}(\vek{q})) = {\cal O}(q^2)$. Collecting the terms 
appearing in $P$,   
the long-wavelength instantaneous 
coupling is then simply given by:
\begin{equation}
R(t) P p(\vek{q}) =  c^2 \rho(\vek{q},t) \, .
\end{equation}
Let's turn now to the retarded couplings. 
From \gl{PiLR}, one has to consider only the
 couplings of $p(\vek{q})$ with $\dot{Q}_{kl}(\vek{q})$ and with 
$J_k(\vek{q})$, and these couplings involve 
$(\ddot{Q}_{kl}(\vek{q}) | R'(t-s) | p(\vek{q}))$ and 
$(\Pi_{kl}(\vek{q}) | R'(t-s) | p(\vek{q}))$. As $\ddot{Q}_{kl}(\vek{q})$ is
traceless while $\Pi_{kl}(\vek{q})$ is not, the first term is 
${\cal O}(q^2)$ while the second is of order unity.  
Consequently, to lowest order in $q$,  
we need to consider only the retarded coupling to $J_k(\vek{q})$. 
It is convenient 
to introduce the long-wavelength pressure correlator in the form: 
\begin{equation}\label{bulk}
(p | R'(t) | p ) \frac{n}{k_B T} = \eta_b(t) \, .
\end{equation}
(Here, and in the rest of the paper, omitting the wave-number dependence 
indicates that
the quantity is to be evaluated for $\vek{q} \to 0$).  

One thus obtains:  
\begin{subequations}
\begin{eqnarray}\label{20a}
& & i q_l (\Pi_{kl}(\vek{q}) | R'(t-s) | p(\vek{q})) \nonumber \\
& & = i q_k \eta_b(t-s) 
k_B T/n + 
{\cal O}(q^3) \, ,
\end{eqnarray}
once  we make use of the fact that 
$(\pi_{kl}(\vek{q}) | R'(t) | p(\vek{q}))$ is of order 
${\cal O}(q^2)$. 
Introducing \gl{20a} into the expression of the retarded coupling
and replacing, in \gl{pressuresub}, $J_k(\vek{q})$ by $m n v_k(\vek{q})$, one
obtains the standard, retarded, constitutive equation for the 
fluctuating pressure: 
\begin{eqnarray}
\label{scalar}
p(\vek{q},t) & = &  c^2 \rho(\vek{q},t) 
+ i \int_0^t \eta_b(t-s) q_k 
v_k(\vek{q},s) ds \nonumber \\
& & + noise \, , 
\end{eqnarray}
which is the Fourier transform of the usual equation:
\begin{eqnarray}
-p(\vek{r},t) & = &  -c^2 \rho(\vek{r},t) 
+  \int_0^t \eta_b(t-s) \mbox{ div } \vek{v}(\vek{r},s) ds \nonumber \\
& & + noise \, .
\end{eqnarray}
\end{subequations}

Let us now apply the same technique to $\pi_{ij}(\vek{q})$, 
the traceless
part of $\Pi_{ij}(\vek{q})$. 
Multiplying \gl{opid} by $\pi_{ij}(\vek{q})$ from the right yields
the Langevin equation for the traceless part of the momentum current tensor. 
In order to evaluate the instantaneous couplings,
one needs to compute 
the static correlations of this tensor
with the distinguished variables. 
Because of \gl{rotinv} and of the time reversal symmetry,
one is left with the sole evaluation of 
$(Q_{kl}(\vek{q}) | \pi_{ij}(\vek{q}))$
for which neither tensorial nor time reversal symmetry constraints apply, in 
the 
long-wavelength limit. 
However from 
momentum conservation:  
\begin{subequations}
\begin{eqnarray}\label{Qpi}
& & i q_j ( Q_{kl}(\vek{q}) | \pi_{ij}(\vek{q}) )
= (Q_{kl}(\vek{q})  | i {\cal L} J_i(\vek{q})) \nonumber \\
& &  = - (\dot{Q}_{kl}(\vek{q}) | J_i(\vek{q}) ) \, ,
\end{eqnarray}
\end{subequations}
where the first equality is correct up to order ${\cal O}(q^2)$ due to the 
traceless character of $Q_{kl}(\vek{q})$. The r.h.s. of \gl{Qpi} is 
equal to zero at every order in $q$, \gl{cross}, so that there is no 
instantaneous coupling of $\pi_{ij}(\vek{q})$ with the distinguished 
variables, in the leading order in $q$ considered in the present paper.
Hence, one is left with the evaluation of the memory kernel,
which splits into two parts: 
 
-the traceless momentum current tensor auto-correlator which, in line with
\gl{bulk}, we write in the form: 
\begin{eqnarray}\label{shearviscosity}
& &
\left( \pi_{kl} | R'(t) | \pi_{ij} \right) \frac{n}{k_B T} 
= \eta_s(t) \Delta_{ij,kl} \, ,
\end{eqnarray}
and we call $\eta_s(t)$ 
 the time-dependent shear 
viscosity, as it couples the 
momentum current to the strain rate, see
 \gl{trans}.

-the coupling of the traceless momentum current 
with the corresponding orientational force, that we write as:
\begin{eqnarray}\label{mu}
& & \left( \ddot{Q}_{kl} | R'(t) | \pi_{ij}\right) 
\frac{1}{\Omega^2} =  -\mu(t) \Delta_{ij,kl} \, .
\end{eqnarray}

For reasons similar to those used in \gl{20a}:
\begin{subequations}
\begin{eqnarray}
& &   i q_l (\Pi_{kl}(\vek{q}) | R'(t-s) | \pi_{ij}(\vek{q})) \\
& & =  i q_l (\pi_{kl}(\vek{q}) | R'(t-s) | \pi_{ij}(\vek{q})) \nonumber \\
& & =  i q_l \eta_s(t-s) \frac{k_B T}{n} \Delta_{kl,ij} \, ,
\end{eqnarray}
and:
\begin{eqnarray}
\label{tau}
& & i q_l \frac{J_k(\vek{q})}{n} \Delta_{kl,ij} = i m [ q_i v_j(\vek{q}) 
+  q_j v_i(\vek{q}) - \frac{2}{3} \delta_{ij} 
q_k v_k(\vek{q}) ] 
\nonumber \\
& & \equiv -m \tau_{ij}(\vek{q}) \, ,
\end{eqnarray}
where $\tau_{ij}(\vek{q})$ is the strain rate tensor, 
so that: 
\begin{eqnarray}
& & R(s) |J_k(\vek{q})) \frac{1}{m k_B T} i q_l (\Pi_{kl}(\vek{q}) | R'(t-s)
| \pi_{ij}(\vek{q})) \nonumber \\
& & 
= - \tau_{ij}(\vek{q},s) \eta_s(t-s) \, .
\end{eqnarray}
Similarly:
\begin{eqnarray}
& & - R(s) 
| \dot{Q}_{kl}(\vek{q}) ) \frac{1}{2 \Omega^2} (\ddot{Q}_{kl}(\vek{q}) | 
R'(t-s) | \pi_{ij}(\vek{q}))  
\nonumber \\
& & = 
R(s) 
| \dot{Q}_{kl}(\vek{q})) \frac{1}{2} \Delta_{kl,ij} \mu(t-s)   \nonumber \\
& & =\dot{Q}_{ij}(\vek{q},s) \mu(t-s) \, . 
\end{eqnarray}
\end{subequations}
Then, the generalised constitutive equation for $\pi_{ij}(\vek{q},t)$ reads:
\begin{eqnarray}
\label{trans}
 \pi_{ij}(\vek{q},t) 
& = &  -\int_0^t \eta_s(t-s) \tau_{ij}(\vek{q},s)  ds \nonumber \\ 
& & + \int_0^t  \mu(t-s) \dot{Q}_{ij}(\vek{q},s)  ds + noise \, .
\end{eqnarray}
Combining  \gl{scalar} and \gl{trans}, one obtains: 
\begin{eqnarray}\label{momentumcurrent}
  \Pi_{ij}(\vek{q},t) 
& = & 
\delta_{ij} c^2 \rho(\vek{q},t) 
+ i \delta_{ij} \int_0^t \eta_b(t-s) q_k v_k(\vek{q},s) ds \nonumber \\
& & 
- \int_0^t \eta_s(t-s) \tau_{ij}(\vek{q},s) ds \nonumber \\ 
& & + \int_0^t  \mu(t-s) \dot{Q}_{ij}(\vek{q},s)  ds +  noise \, ,
\end{eqnarray}
which is exactly the Fourier transform of Eq. (3) of Part I, once one has 
noted that $\Pi_{ij}(\vek{q},t)$ is the opposite of the stress
tensor, $\sigma_{ij}(\vek{q},t)$. 

To derive an equation of motion  for $\ddot{Q}_{ij}(\vek{q})$, 
we again make use of  \gl{opid}.  
For the instantaneous contribution,
 $R(t) P  \ddot{Q}_{ij}(\vek{q})$, only the term involving 
the orientation in the projector $P$ needs to be considered, the other
terms dropping out for tensorial or time-reversal symmetry considerations. 
Since: 
\begin{subequations}
\begin{eqnarray}
(Q_{kl}(\vek{q}) | \ddot{Q}_{ij}(\vek{q})) & = &  
- (\dot{Q}_{kl}(\vek{q}) |  \dot{Q}_{ij}(\vek{q})) \nonumber \\
& = &  - \Omega^2 \Delta_{kl,ij}\, , 
\end{eqnarray}
\begin{equation} 
R(t) P \ddot{Q}_{ij}(\vek{q}) = - \omega_0^2 Q_{ij}(\vek{q},t) \, ,
\end{equation}
with the axial libration frequency:  
\begin{equation}\label{omega}
\omega_0 = \Omega/S \, .
\end{equation}
\end{subequations} 
The evaluation of the retarded couplings proceeds along
 the same lines as for the
momentum current tensor. Defining:
\begin{subequations}
\begin{eqnarray}\label{Gamma}
( \ddot{Q}_{kl} | R'(t) | \ddot{Q}_{ij}) \frac{1}{\Omega^2 }
& = & 
\Gamma'(t) \Delta_{kl,ij} \, ,
\end{eqnarray}
and: 
\begin{equation}\label{lambda}
\Lambda' = \frac{\Omega^2 n}{k_B T} = \frac{2n}{5 I} \, ,
\end{equation}
one easily obtains, with the help of Eqs. (\ref{PiLR}), and
(\ref{tau}):
\begin{eqnarray}
P i {\cal L} R'(t-s) \ddot{Q}_{ij}(\vek{q}) & = &  
- | \dot{Q}_{ij}(\vek{q})) \Gamma'(t-s)  \nonumber \\
& & 
+ \Lambda' |\tau_{ij}(\vek{q})) \mu(t-s) \, ,
\end{eqnarray}
once one has noted that, because of the traceless character of 
$\ddot{Q}_{ij}(\vek{q})$:
\begin{equation}
(\Pi_{kl} | R'(t-s) | \ddot{Q}_{ij}) 
= (\pi_{kl} | R'(t-s) | \ddot{Q}_{ij})  \, .
\end{equation}
\end{subequations}
Collecting the various terms, one thus obtains:
\begin{eqnarray}
\label{osci}
 \ddot{Q}_{ij}(\vek{q},t)  & = &   - \omega_0^2 Q_{ij}(\vek{q},t) 
- \int_0^t \Gamma'(t-s) \dot{Q}_{ij}(\vek{q},s) ds \nonumber \\
& &  
+ \Lambda' \int_0^t \mu(t-s) \tau_{ij}(\vek{q},s) ds 
\nonumber \\
& & + noise \, .
\end{eqnarray}
The same memory kernel, $\mu(t)$, occurs in the constitutive equation for
the orientational force, \gl{osci}, as response to a momentum gradient,
and in the equation for the 
 momentum current, \gl{trans}, as a reaction to an orientational current. 
This can be considered as a general consequence of  Onsager's principle, 
and it appears, here, naturally  
as the result of
 the use of the Zwanzig-Mori formalism.  
Equation (\ref{osci}) is, as expected, the Fourier transform of Eq. (4)
 of Part 
I, as  briefly argued in \cite{Latz2001}. 
The Zwanzig-Mori formalism thus leads to the microscopic derivation of the
equations proposed, on a phenomenological basis, in \cite{Drey1,Drey2}.
There are, nevertheless, already two
bonuses. One is the precise definitions of 
$\Lambda'$, \gl{lambda}, in terms of quantities a priori known, and 
of $\omega_0$, \gl{omega}, which can be obtained from thermal averages
of $(Q_{ij} | Q_{kl})$ and 
$(\dot{Q}_{ij} | \dot{Q}_{kl})$. The second bonus
is the precise definitions, through $R'(t)$ of the four memory kernels,
$\eta_b(t), \eta_s(t), \mu(t)$ and $\Gamma'(t)$, Eqs. 
(\ref{bulk},\ref{shearviscosity},\ref{mu}) and (\ref{Gamma}). We shall show, 
in the next two Sections, that these expressions allow:

- on the one hand (Section III) to precisely define under which conditions,
all the Brillouin intensities, derived or recalled in Part I, are positive 
whatever the frequency, within the scattering model used in 
\cite{Dreyfus2002a}.

- on the other hand (Section IV) to show, through Kubo's formulae,
 that these
kernels can, directly or indirectly depending on which one is considered, 
be measured as correlation functions of $q\to 0$ dynamical variables. 

\section{The Onsager Relations and the positiveness of the spectra}
\subsection{Summary of the light scattering results of Part I}
In Part I, \cite{Dreyfus2002a}, making use of the equations of motions 
(Eqs. (\ref{massconservation}), (\ref{momentumconservation}), 
(\ref{momentumcurrent}) and (\ref{osci})), 
we gave an expression for 
the intensity of the VV light-scattering spectrum under the assumption that
the fluctuations 
of the dielectric tensor could be written as the linear combination:
\begin{equation}
\label{scattering_model}
\epsilon_{ij}(\vek{q}) = a \delta_{ij} \rho(\vek{q}) 
+ b Q_{ij}(\vek{q}) \, .
\end{equation}
Taking the convention that the Laplace transform of $f(t)$ would be 
$f(\omega) = LT[f(t)](\omega) = i \int_0^\infty dt f(t) \exp(-i \omega t)$, 
this intensity was expressed (see Eq. 36, Part I) in terms of all the 
quantities defined in Section II, and of 
$\langle |Q_{\perp \perp'}^0 |^2 \rangle = S^2 $, see \gl{S}. 
Using Eqs. (\ref{omega}) and (\ref{lambda}) 
which relate
$\omega_0, \Lambda', \Omega$ and $S$, the result obtained in Part I 
can be cast into the form 
\footnote{In this Section, we neglect in the expression of the
intensities, the $\delta(\omega)$ terms related to the $Im (1/\omega)$ 
contributions.}:
\begin{subequations}
\begin{eqnarray}\label{PVV}
\lefteqn{ I_{VV}(\vek{q},\omega) =   \frac{1}{\omega} Im \Bigg\{ 
-\frac{4 b^2}{3}  \frac{ \Omega^2}{D(\omega)}  }
\nonumber \\
 & & 
+ q^2 \left[ a + \frac{2b \Lambda'}{3 m n}   r(\omega) \right]^2 
 m^2 v^2 P_L(q,\omega) \Bigg\} \, ,
\end{eqnarray}
where $P_L(q,\omega)$ is the longitudinal phonon propagator:
\begin{eqnarray}
P_L(q,\omega)^{-1} = \omega^2- q^2 c^2 - q^2 \omega \eta_L(\omega) /mn \, , 
\end{eqnarray}
with:
\begin{eqnarray}\label{defetaT}
\eta_L(\omega) &= &  \eta_b(\omega) + \frac{4}{3} [ \eta_s(\omega) 
- \frac{\Lambda'}{\omega} D(\omega) r(\omega)^2 ] \nonumber \\
&  \equiv & 
k_L(\omega) - \frac{4}{3} \frac{\Lambda'}{\omega} D(\omega) r(\omega)^2 
\nonumber \\
& \equiv & \eta_b(\omega) + \frac{4}{3} \eta_T(\omega) \, ,
\end{eqnarray}
\end{subequations}
\begin{equation}\label{Dom}
D(\omega) = \omega_0^2 + \omega \Gamma'(\omega) -\omega^2 \, , 
\end{equation}
\begin{equation}
\label{Pockels1}
r(\omega) = \frac{\omega \mu(\omega)}{ D(\omega)} \, .
\end{equation}
$\eta_T(t)$, defined through the last line of \gl{defetaT} is what we shall
call the transverse viscosity.

Similarly, the expression for the intensity of the VH light-scattering 
spectrum, already derived in \cite{Drey1,Drey2} within the same model, 
was recalled in Part I (Eq. (48)); with
 the present notations, it reads:
\begin{subequations}
\begin{eqnarray}\label{PVH}
\lefteqn{  
I_{VH}(\vek{q},\omega)  =   \frac{b^2}{\omega} Im \Bigg\{
\frac{- \Omega^2}{D(\omega)} }
\nonumber \\
& & + q^2 \left[ \frac{\Lambda' r(\omega)}{ m n}
\right]^2 \cos^2 \frac{\theta}{2} m^2 v^2 P_T(q,\omega) \Bigg\} \, ,
\end{eqnarray}
where $\theta$ is the scattering angle and:
\begin{equation}\label{PT}
P_T(q,\omega)^{-1} = \omega^2 - q^2 \omega \eta_T(\omega)/mn
\end{equation}
is the transverse phonon propagator. It is convenient to separate out the 
angular contribution in \gl{PVH} by rewriting it in the form:
\begin{equation}\label{I1VH}
I_{VH}(\vek{q},\omega) = b^2 \left[ \sin^2 \frac{\theta}{2}  I_{BD}(\omega)
+ \cos^2 \frac{\theta}{2} 
I_T(q,\omega) \right] \, ,
\end{equation}
with:
\begin{equation}\label{I_D}
I_{BD}(\omega) = \frac{1}{\omega} Im \left[ \frac{- \Omega^2}{D(\omega)} 
\right] \, ,
\end{equation}
\begin{eqnarray}\label{I_T}
I_T(q,\omega) &= & \frac{1}{\omega} Im \Bigg\{ \frac{- \Omega^2}{D(\omega)} 
\nonumber \\ 
& & +
 q^2 m^2 v^2  \left[ \frac{\Lambda' r(\omega)}{mn} \right]^2 
P_T(q,\omega)
\Bigg\} \, .
\end{eqnarray}
\end{subequations}

\subsection{Necessary conditions on the memory kernels}
In the present part of Section III, 
we show some general properties of the four memory kernels 
$\eta_b(t), \eta_s(t), \mu(t)$ and $\Gamma'(t)$ that can be derived from their
microscopic expressions. These properties
 are of interest for the light scattering spectra and, in particular, 
they are such  that, when fulfilled,
 the spectra are positive whatever the frequency and
the ratio $b/a$ of \gl{scattering_model}.

Firstly,
from their  definitions, Eqs. 
(\ref{bulk}, \ref{shearviscosity}, \ref{mu}, \ref{Gamma}), 
one checks that the memory kernels are real and have even time parity. 
Furthermore, by taking the special linear combinations:
\begin{subequations}
\begin{eqnarray}
\pi_{20} &= & [2 \pi_{zz} - \pi_{xx} 
 - \pi_{yy} ]/\sqrt{12} \, , \\
Q_0 &= & [2 Q_{zz} - Q_{xx} 
 - Q_{yy} ]/\sqrt{12} \, ,
\end{eqnarray}
\end{subequations}
$\eta_s(t)$ and $\Gamma'(t)$ can be written as auto-correlation functions 
similar to $\eta_b(t)$:
\begin{subequations}
\begin{eqnarray}\label{transcorr}
\eta_s(t) & = & \frac{n}{k_B T} 
(\pi_{20} | R'(t) | \pi_{20}) \, ,
\\
\Gamma'(t) & = & \frac{1}{\Omega^2}
(\ddot{Q}_{0} | R'(t) | \ddot{Q}_{0}) 
\, .
\end{eqnarray}
\end{subequations}
Thus the Laplace-transforms of the memory kernels have the usual properties
described, e.g., in 
\cite{Forster75,Berne76,hansen} (see also Appendix C). In particular,
this Appendix shows that these Laplace transforms 
 are analytic in the lower complex half plane and that the 
inequalities: 
\begin{subequations}\label{positive}
\begin{equation}\label{Imetab}
Im \, \eta_b(\omega) \geq 0 \, ,
\end{equation}
\begin{equation}\label{Imetas}
 Im \, \eta_s(\omega) \geq 0 \, ,
\end{equation}
\begin{equation}\label{ImGamma}
Im \, \Gamma'(\omega) \geq 0 \, ,
\end{equation}
\end{subequations} 
hold for 
all complex $\omega$ with $Im \, \omega < 0$. 

The mixed correlation function: 
\begin{subequations}
\begin{equation}
- \Omega^2 \mu(t) = ( \ddot{Q}_0 | R'(t) | \pi_{20} ) \, ,  
\end{equation}
can be read as an off-diagonal element of the matrix correlator built on
$ \ddot{Q}_0$ and $\pi_{20}$. Since 
the imaginary part of the Laplace transform of this matrix is
 positive semidefinite (see Appendix C, \gl{Onsager}), one obtains, with the
help of \gl{lambda}, the inequality: 
\begin{eqnarray}\label{Cauchy}
[Im \eta_s(\omega)] 
[Im \Gamma'(\omega)] - \Lambda' [Im \mu(\omega) ]^2 \geq 0 \, ,
\end{eqnarray}
for all $\omega$ in the lower complex half- plane. 
The system of inequalities Eqs. 
(\ref{positive},
\ref{Cauchy}) 
is a generalisation of Onsager's relations to finite 
frequencies:  from the microscopic approach, one
 obtains  
that the imaginary part of the matrix of kinetic coefficients is
positive definite for any frequency. 

Let us mention one useful consequence. 
First by Fourier back-transform: 
\begin{eqnarray}
\Lambda' \mu(t)^2 &= & \Lambda' \left[ \int \frac{d\omega}{\pi} 
\cos(\omega t)
Im \mu(\omega) \right]^2 \nonumber \\
& \leq & \Lambda' \left[ \int \frac{d\omega}{\pi} 
\left| Im \mu(\omega) \right| \right]^2 \nonumber \\
& \leq &  \left[ \int \! \! \frac{d\omega}{\pi}  Im \Gamma'(\omega)  \right]
\left[ \int \! \! \frac{d\omega}{\pi}  Im \, \eta_s(\omega)  \right] \, ,
\end{eqnarray}
where in the last line we used Eq. (\ref{Cauchy}).  
The last relation implies that  the 
translation-rotation coupling is bounded by:
\begin{equation}
\Lambda' \mu(t)^2  \leq  \eta_s(t=0) \Gamma'(t=0) \, .
\end{equation}
\end{subequations}

\subsection{Positiveness of the light scattering spectra and further relations}
The four  inequalities, Eqs. (\ref{positive},\ref{Cauchy}),
turn out to be  sufficient to prove that
the light-scattering spectra, Eqs. (\ref{PVV}) and (\ref{PVH}), 
are positive for 
any frequency. The proof 
will  be given for real frequencies only since 
the algebra greatly simplifies. By similar methods,
 one can extend the proof
to hold for all frequencies in the lower complex half-plane.
Let us first recall that if ${\bf A}$ is a symmetric complex matrix,
then one can  write
$ Im ({\bf A}^{-1}) = -[ 
Im {\bf  A} + (Re {\bf A}) (Im {\bf A})^{-1} (Re {\bf A})]^{-1}$. 
If $Im {\bf A}$ is a positive definite matrix, one  proves that 
$(Re {\bf A}) (Im {\bf A})^{-1} (Re {\bf A})$ has the same property, so that 
$-Im ({\bf A}^{-1})$ is also a positive definite matrix;
in particular, its 
diagonal elements are positive
\footnote{One easily includes the 
case where $Im {\bf A}$ is not invertible due to a vanishing eigenvalue 
by adding an arbitrarily small imaginary multiple of the unit matrix; 
this simply changes the strict inequalities into weak ones. 
In order to avoid
this complication,
 we shall use Eqs. 
(\ref{positive},
 \ref{Cauchy})
 as strict inequalities.}.
Let us  now make use of this property to prove the positiveness of 
the spectra, starting with the VH spectrum written as \gl{I1VH}. 

First:
\begin{subequations}
\begin{eqnarray} 
I_{BD}(\omega) = 
\frac{ \omega_0^2 Im \Gamma'(\omega)}{ [\omega_0^2 - \omega^2 + \omega Re \Gamma'(\omega)]^2 + [\omega Im \Gamma'(\omega) ]^2 } \, ,
\end{eqnarray}
so that $I_{BD}(\omega)$ is always positive. Second, consider the matrix: 
\begin{eqnarray}\label{transmatrix}
& & \left[
\begin{array}{cc}
F_T(q,\omega)/\Omega^2* & * \\
* & \omega^2 C_T(q,\omega)/(m^2 v^2 q^2)
\end{array}
\right]
= \nonumber \\
& & -  
\left[
\begin{array}{ccc}
\omega D(\omega) & q \Omega \omega \mu(\omega)/(mv)  \\
q \Omega \omega \mu(\omega)/(mv) & -\omega + q^2\eta_s(\omega)/mn  
\end{array}
\right]^{-1} \, .
\end{eqnarray}
(here the matrix elements we are not interested in are abbreviated by 
asterisks). 
One checks that, 
for real $\omega$, 
the imaginary part of the matrix on the r.h.s is positive definite: indeed,
 its 
diagonal elements are positive (Eqs. \ref{Imetas}, \ref{ImGamma}) 
while the corresponding 
$2\times 2$ determinant is proportional to the l.h.s. of \gl{Cauchy}. 
Consequently, so does the imaginary part of the l.h.s. of \gl{transmatrix}. 
In particular, the diagonal
elements on the left-hand side are positive.
After some algebra one finds that:
\begin{eqnarray}
Im \, F_T(q,\omega) & = &  I_T(q,\omega) \, ,
\end{eqnarray}
\begin{eqnarray}
Im \, C_T(q,\omega) &= & \frac{q^2 m^2 v^2}{\omega} Im \, P_T(q,\omega)  \, .
\end{eqnarray}

As both $I_{BD}(\omega)$ and $I_T(q,\omega)$ are positive whatever $\omega$ real,
the depolarised light-scattering spectrum, \gl{I1VH} is always positive. 
Also, from the form of $P_T(q,\omega)$, \gl{PT}, the sign of $Im P_T(q,\omega)$
is the same as that of $\eta_T(\omega)$; this implies: 
\begin{equation}\label{Imetat}
Im \, \eta_T(\omega) \geq 0 \, .
\end{equation}
\end{subequations}
Thus, 
in spite of its intricate expression, Eq. (\ref{defetaT}),
one can  prove that $Im \eta_T(\omega)$ is always positive, a result
which will  be obtained again through the 
Green-Kubo technique in Section IV. 

For the polarised spectrum, let us consider the similar matrix:
\begin{subequations}
\begin{eqnarray}\label{longi}
& & \left[
\begin{array}{cc}
 F_L(q,\omega)/ \Omega^2 
& \omega G_L(q,\omega)/(\Omega m v q) \\
\omega G_L(q,\omega)/(\Omega m v q) & \omega^2 C_L(q,\omega)/(m^2 v^2 q^2)
\end{array}
\right]
= \nonumber \\
& & -  
\left[
\begin{array}{ccc}
\omega D(\omega) & \sqrt{4/3} q 
\frac{\displaystyle \Omega \omega \mu(\omega)}{\displaystyle mv}  \\
& & \\
\sqrt{4/3} q \frac{\displaystyle \Omega \omega \mu(\omega)}{\displaystyle mv} 
&  \frac{\displaystyle c^2 q^2-\omega^2}{\displaystyle \omega} 
+  \frac{\displaystyle q^2 k_L(\omega)}{\displaystyle mn}   
\end{array}
\right]^{-1} \nonumber \, .\\
\end{eqnarray}
Again, the imaginary part of the matrix on the r.h.s of \gl{longi} is positive
definite; its diagonal elements are positive, 
Eqs. (\ref{positive}),  
while the corresponding
$2\times 2$ determinant is proportional to:
\begin{equation}
[Im \, \Gamma'(\omega)] 
[ Im \, \eta_b(\omega) + \frac{4}{3} Im \, \eta_s(\omega) ]
- \frac{4}{3} \Lambda' [Im \, \mu(\omega)]^2 \, .
\end{equation}
This term is also positive, because 
of Eqs. (\ref{Cauchy}) and (\ref{Imetab},\ref{ImGamma}). 
Explicit evaluation of the
inverse of the matrix on the r.h.s. of \gl{longi}
yields:
\begin{eqnarray}
Im F_L(q,\omega) & = & \frac{3}{\omega} Im 
\Big\{ \frac{-\Omega^2}{3 D(\omega)} \nonumber \\
& &   
+ q^2  \left[ \frac{2}{3}\frac{\Lambda' r(\omega)}{m n} \right]^2 m^2 v^2 
P_L(q,\omega) \Big\} \, ,
\end{eqnarray} 
\begin{equation}
Im G_L(q,\omega) = \frac{\sqrt{3}}{\omega} q^2 m^2 v^2  Im \left\{ 
 \frac{2}{3} \frac{\Lambda' r(\omega)}{m n} 
 P_L(q,\omega) \right\} \, ,
\end{equation}
\begin{equation}\label{ImCL}
Im C_L(q,\omega) = \frac{q^2 m^2 v^2  }{\omega} Im 
P_L(q,\omega) \, .
\end{equation}

Since the imaginary part of the 
matrix on the l.h.s. of  \gl{longi} is positive definite, 
this is also true for the matrix whose elements are 
$F_L(q,\omega), G_L(q,\omega)$ and $C_L(q,\omega)$. In consequence:
\begin{eqnarray}
I_L(q,\omega) &= &  
a^2 Im C_L(q,\omega) 
+ \frac{2}{\sqrt{3}} a b \, Im \, G_L(q,\omega) \nonumber \\
&  & 
+ \frac{b^2}{3} Im F_L(q,\omega) \geq 0 \, .
\end{eqnarray}
Since 
\begin{equation}
I_{VV}(q,\omega) = b^2 I_{BD}(\omega) + I_L(q,\omega) \, , 
\end{equation}
\end{subequations}
the VV spectrum, \gl{PVV}, is positive, whatever $a$ and $b$. [Note that the
same technique could be applied to the HH spectrum, Eq. (43) of Part I, to
prove that it is also positive, whatever the scattering angle.]
Equation (\ref{ImCL})
 can also be used to prove that $Im \, \eta_L(\omega) > 0$, 
a conclusion which already resulted from Eqs. (\ref{Imetat}) 
and (\ref{Imetab}). 
Let us stress that the positiveness of $I_{VV}(q,\omega)$ for any 
$\omega$, whatever $q$, is not a trivial result as $I_L(q,\omega)$ is the
sum of a $q$-independent term, proportional to $I_{BD}(\omega)$,
and of a $q$-dependent term. One could naively think that $I_{VV}(q,\omega)$
could be always
 positive only if the same would be true for this
$q$-dependent term. Figure 3 of Part I, \cite{Dreyfus2002a} shows
that this is not the case. In fact, though Eq. (\ref{Cauchy}) does not invoke 
$q$, it insures that $Im \Gamma'(\omega)$ is large enough, whatever $\omega$,
for the sum of the two terms of $I_L(q,\omega)$ to be positive,
independently of the value of $q$. A similar argument holds for the
$I_{VH}(q,\omega)$ spectrum.

\section{The Green-Kubo Approach to the memory kernels}
\subsection{Preliminary Remarks}
The comparison between measured Brillouin spectra and their predicted
intensities (Eqs. \ref{PVV} and \ref{PVH}) 
requires the knowledge of the four memory 
kernels $\eta_b(t), \eta_s(t), \mu(t)$ and $\Gamma'(t)$. 
Although microscopically well defined, those cannot be evaluated exactly, 
so that in practice, they
are frequently taken as empirical fit functions.
Whereas the correct memory kernels are guaranteed to reflect
all the restrictions of the correlated motion of the translational
and orientational degrees of freedom, i.e. automatically fulfill the
relations, Eqs. (\ref{Imetab},\ref{Imetas},\ref{ImGamma}) and (\ref{Cauchy}), 
this needs not  be true for these
empirical functions. Hence, one has to carefully choose their 
parameters so that these relations are fulfilled.

A possible intermediate approach consists in obtaining 
information on those memory kernels through MD calculations of some
realistic model of the supercooled liquid under consideration.
Because the memory kernels are $q\to 0$ limits of correlation functions 
of specific variables, they can, in principle, be computed
from these MD calculations. Yet, 
these kernels, Eqs. 
(\ref{bulk},\ref{shearviscosity},\ref{mu}) and (\ref{Gamma}), 
are written in terms of the
reduced operator $R'(t)$. As the latter has no  easy formulation,
this apparently reduces drastically the value of the preceeding remark.
We show,
in the present Section,
that the microscopic approach of Section II
allows for the determination of expressions of $\eta_b(t)$, and $\eta_T(t)$,
the bulk and
transverse viscosities, which coincide with the usual Green-Kubo formulae:
they can be directly determined as the correlation functions of 
 variables accessible
 in a MD calculation.

Conversely, no such direct determination of $\Gamma'(t)$ and $\mu(t)$ is 
possible; 
 their Laplace transforms, 
can be obtained through the computation of the Laplace 
transforms of the correlation functions of  other dynamical
variables but the determination of $\Gamma'(\omega)$ and $\mu(\omega)$ will
be rather indirect, as we shall see; for technical reasons, we shall
start
with this second aspect and will turn, later on, to the 
determination of the viscosity kernels.

Let us start by recalling that, in \gl{osci}, the 'noise term' is 
equal to $R'(t) \ddot{Q}_{ij}(\vek{q})$ so that one can rewrite this 
equation as:
\begin{subequations}
\begin{eqnarray}\label{RddQ}
 R'(t) \ddot{Q}_0(\vek{q})  & = &  
 \ddot{Q}_0(\vek{q},t) + \omega_0^2 Q_0(\vek{q},t) \nonumber \\
& & 
+ \int_0^t \Gamma'(t-s) \dot{Q}_0(\vek{q},s) ds \nonumber \\ 
& &
- \Lambda' \int_0^t \mu(t-s)
\tau_0(\vek{q},s) ds \, ,
\end{eqnarray}
with:
\begin{equation}
\tau_0(\vek{q}) = [2 \tau_{zz}(\vek{q}) - \tau_{xx}(\vek{q}) 
-\tau_{yy}(\vek{q})]/\sqrt{12} \, .
\end{equation}
\end{subequations}
When computing $\mu(t)$ or $\Gamma'(t)$ through Eqs. (\ref{mu},\ref{Gamma}), 
the $\vek{q}
\to 0$ limit is taken, and $\tau_0(\vek{q})$ is ${\cal O}(q)$ (see \gl{tau});
the last term of \gl{RddQ} may thus be dropped. Multiplying 
both sides of \gl{RddQ}, from the left, by $\ddot{Q}_0$, performing a thermal
average and a Laplace transform yields:
\begin{subequations}
\begin{eqnarray}\label{OmGamma}
-\Omega^2 \Gamma'(\omega) &= &  LT[ (\ddot{Q}_0 | R'(t) | \ddot{Q}_0 )](\omega)
\nonumber \\
&= &  (\omega_0^2 + \omega \Gamma'(\omega) - \omega^2) LT[( Q_0(t) | \ddot{Q}_0)](\omega) \nonumber \\
& & 
+ [\omega - \Gamma'(\omega)] (Q_0 | \ddot{Q}_0 ) \, .
\end{eqnarray}
As $(\ddot{Q}_0 | Q_0 ) = - (\dot{Q}_0 | \dot{Q}_0 ) = - \Omega^2$, the 
$\Omega^2 \Gamma'(\omega)$ drops out of \gl{OmGamma}: $\Gamma'(\omega)$ 
is not directly determined by considering the 'noise term': 
its indirect determination is nevertheless
possible through Eq. (\ref{OmGamma}) as:
\begin{equation}\label{QddQ}
LT[ (Q_0(t) | \ddot{Q} )](\omega) = \frac{\omega \Omega^2}{D(\omega)} \, .
\end{equation}
The l.h.s. of \gl{QddQ} can be obtained from the correlation of $Q_0(t)$ with
$\ddot{Q}_0$. Nevertheless, it is simpler to write:
\begin{eqnarray}\label{QddQ2}
& &LT[ (Q_0(t) | \ddot{Q}_0)](\omega) = 
LT[ (\ddot{Q}_0(t) | Q_0)](\omega) 
\nonumber \\
& & = 
-\omega^2 LT[ (Q_0(t) | Q_0)](\omega) + \omega (Q_0 | Q_0) \, .
\end{eqnarray}
Eqs. (\ref{QddQ},\ref{QddQ2}) can be recast into the form:
\begin{equation}\label{Q0Q0}
LT[ (Q_0(t) | Q_0)](\omega) 
= \frac{S^2}{\omega} \left[ 1 - \frac{\omega_0^2}{D(\omega)} \right]\, .
\end{equation}
\end{subequations}
The fancy technique we have just used simply recovers Eq. (34) of Part I 
which was directly obtained from the phenomenological  equations of 
motion. The latter have been microscopically derived in Section II, and
this derivation implied the neglect of the noise term term $R'(t)$. 
The above given 
proof of \gl{Q0Q0} can be considered as a consistency check for  the 
use of the 'noise term' to derive valuable results, a 
technique we shall now use to derive useful expressions for
$\eta_b(t)$ and $\eta_T(t)$.

Before doing that, let us multiply the $q\to 0$ limit of \gl{RddQ}, on the
left, by $\pi_{20}$. Performing similar manipulations as above, one obtains:
\begin{subequations}
\begin{eqnarray}
 -\Omega^2 \mu(\omega) &= & LT[ (\pi_{20} | R'(t) | \ddot{Q}_0)](\omega) 
\nonumber \\
 & = &
D(\omega) LT[ (Q_0(t) | \pi_{20} ) ](\omega) \nonumber \\
& &  -
[\omega- \Gamma'(\omega)] (\pi_{20} | Q_0 ) \, ,
\end{eqnarray}
where $(\pi_{20} | Q_0)$  is equal to zero, due to 
Eqs. (\ref{Qpi}) and (\ref{cross}).
This yields:
\begin{equation}\label{muD}
\mu(\omega) = 
-\frac{D(\omega)}{\Omega^2} LT[ (Q_0(t) | \pi_{20})](\omega) \, ,
\end{equation}
or, equivalently:
\begin{equation}\label{romega}
r(\omega) = - \frac{\omega}{\Omega^2} LT[ (Q_0(t) | \pi_{20} )](\omega) \, .
\end{equation}
\end{subequations}
It should be noted that $r(\omega)/\omega$ 
is of order ${\cal O}(\omega^{-3})$ for 
frequencies $\omega \gg \omega_0$; this  is 
consistent with the 'sum rule' associated with  
$(Q_0 | \pi_{20} ) = 0$. $r(\omega)$ is the function which 
couples the (longitudinal and transverse) phonon propagator to the
light scattering mechanism via the orientational part of these
excitations, see Eqs. (\ref{PVV}) and (\ref{PVH}). Equation (\ref{romega})
shows that $r(t)$ can be directly obtained as the time derivative of the 
correlation function of $Q_0(t)$ with $\pi_{20}$, but that $\mu(t)$ is 
not directly accessible; it can be obtained
only once $r(\omega)$ and $D(\omega)^{-1}$ have been determined by the 
MD calculation.

\subsection{Expressions of $\eta_T(t)$ and $\eta_b(t)$ as 
time correlation functions}
Let  us now use the same 'noise term' technique to express $\eta_T(t)$
and $\eta_b(t)$ as auto-correlation functions of some dynamical 
variables. The 'noise term' of  \gl{trans} is equal to 
$R'(t) \pi_{20}(\vek{q})$, and, in the same $q\to 0$ limit, this equation
simplifies into:
\begin{subequations}
\begin{eqnarray}
R'(t) \pi_{20} = \pi_{20}(t) - \int_0^t \mu(t-s) \dot{Q}_0(s) ds \, .
\end{eqnarray}
Multiplying this equation from the left by $\pi_{20}$, and performing 
the same manipulations as before yields, with the help 
of \gl{shearviscosity}:
\begin{eqnarray}\label{etaD}
\frac{k_B T}{n} \eta_s(\omega) & = &
LT[ (\pi_{20}(t) | \pi_{20} )](\omega) \nonumber \\
& & 
- \mu(\omega) \{ \omega LT[ (Q_0(t) | \pi_{20} )](\omega) - (Q_0 | \pi_{20})
\} \, . \nonumber \\
\end{eqnarray}
Using Eqs. (\ref{muD}), (\ref{Qpi}) and (\ref{cross}), \gl{etaD}
 transforms into:
\begin{equation}
 \eta_s(\omega) = \frac{n}{k_B T} LT[(\pi_{20}(t) | \pi_{20})](\omega) 
+\frac{\Lambda'}{\omega} \frac{\left[\omega \mu(\omega)\right]^2}{D(\omega)} 
 \, .
\end{equation}
From the definition of $\eta_T(t)$, Eq. (\ref{defetaT}),
this equation reads:
\begin{equation}\label{etaT}
\eta_T(t) = \frac{n}{k_B T} (\pi_{20}(t) | \pi_{20} ) \, .
\end{equation}
\end{subequations}
Equation (\ref{etaT}) is the link between the usual Navier-Stokes approach to
the dynamics of supercooled liquids and the most sophisticated 
approach of the present series of papers, which takes explicitly into
account the rotational motion of the molecules (the usual approach is 
recovered by formally putting $\mu(t) \equiv 0$). 
$\eta_T(\omega)$ (see \gl{PT})  is the memory function which governs the 
transverse phonon propagator: 
within
the Green-Kubo formalism of \gl{etaT}, this transverse viscosity is 
proportional to the correlation of the traceless part of the stress tensor, 
$\pi_{20}$, independently of the existence of a rotation-translation coupling.
In other words, the pure center-of-mass viscosity, $\eta_s(t)$, is not
the quantity directly measured by the correlation function of $\pi_{20}$:
$\eta_s(t)$ must be deduced from the simultaneous determination of
$\eta_T(\omega), r(\omega)$ and $D(\omega)$, quantities which can all
be obtained, at least in principle, 
 as correlation functions of some properly chosen 
variables, as we have just shown. Equation (\ref{etaT}) also proves directly,
see Eq. (\ref{Imetat}), that $Im \eta_T(\omega)$ is always
positive, being the Fourier transform of a auto-correlation function.

The same type of technique can be used to determine 
$\eta_b(t)$. In the $q\to 0$ limit, \gl{scalar} reads:
\begin{subequations}
\begin{equation}
R'(t) p = p(t) - c^2 \rho(t) = R(t) [ p - \rho ( \rho | p)/(\rho | \rho) ]
 \, ,
\end{equation}
once Eqs. (\ref{rhocorr}) and (\ref{prhocorr}) have been taken into account, 
or:
\begin{equation}\label{Qnp}
R'(t)p = R(t) Q_n p \, .
\end{equation}
Equation (\ref{Qnp})
 introduces the variable $Q_n p$, which is the part of the pressure
which is orthogonal to the density.
Because of the existence of a $Q$ operator on 
the left hand side of $R'(t)$ (see Appendix B)
which projects out the $\rho$ variable:
\begin{equation}
(Q_n p | R'(t) = ( p | R'(t) \, .
\end{equation}
Thus, \gl{bulk} can be written as:
\begin{equation}\label{GKbulk}
\eta_b(t) = \frac{n}{k_B T} (Q_n p(t) | Q_n p) \, .
\end{equation}
\end{subequations}
Equation (\ref{GKbulk})
 is the analog of the 
 Green-Kubo formulation of the bulk viscosity within the usual
Navier-Stokes formalism: as the rotation-translation coupling does not play 
a role in the bulk viscosity, this usual formulation remains exact in the
more sophisticated present approach. 

Here a comment is in order. Since we did not deal with energy 
fluctuations in the projector, 
the
correlation function $\eta_b(t)$ decays to a non-zero constant  
even for times much larger than the structural relaxation time.  
It is therefore convenient to define a new correlation 
function $\tilde{\eta}_b(t)$ that vanishes at long times by an appropriate
subtraction. One can work  
out the constant from thermodynamic considerations and find:
\begin{eqnarray}
\eta_b(t) = \tilde{\eta}_b(t) + m n (\tilde{c}^2 - c^2) \, ,
\end{eqnarray}
where $\tilde{c}$ is  the adiabatic sound velocity. 
For the Laplace transforms this implies the relation:
\begin{equation}
\omega \eta_b(\omega) = \omega \tilde{\eta}_b(\omega) + mn (\tilde{c}^2 -c^2)
\, .
\end{equation} 
In all correlation functions considered so far, the bulk viscosity appeared
only via the longitudinal phonon propagator. Using the preceeding 
equation $P_L(q,\omega)$ reads:
\begin{equation}
 P_L(q,\omega)^{-1} = 
\omega^2 - \tilde{c}^2 q^2 - \frac{q^2 \omega}{mn} 
[\tilde{\eta}_b(\omega) + \frac{4}{3} \eta_T(\omega)] \, ,
\end{equation}
which shows that the adiabatic sound velocity governs the propagation
of longitudinal phonons. To simplify notations, in the remaining 
part of the paper, we shall  drop the tilde
again and treat $c$ as the adiabatic sound velocity and $\eta_b(t)$ 
as decaying to zero for long times. 

\section{Comparison with previous theoretical approaches}

\subsection{Introduction}
The discussions performed in \cite{Drey2} and \cite{Drey1}
have made clear that the set of Eqs. (\ref{momentumcurrent}) 
and (\ref{osci}) are convenient 
tools to describe the light scattering spectra of molecular liquids, in their
normal and in  their supercooled states, when
 those equations are supplemented
by the dielectric model of Eq. (\ref{scattering_model}). Indeed as soon as the 
four memory functions $\eta_b(t), \eta_s(t), \mu(t)$ and 
$\Gamma'(t)$ are mimicked by reasonably decreasing functions (characterised,
inter alia, by relaxation times, $\tau$, that increase with decreasing 
temperature) the most characteristic features of the VH spectra can be 
described: 

- The back scattering spectrum is mostly characterised by a 
broad high-frequency libration mode, in the vicinity
 of a frequency $\omega_0/2\pi$, and by 
a low-frequency central mode, the line
width of which decreases upon cooling. Both features can be 
approximately reproduced 
by Eq. (\ref{I_D}) with the help of the expression of  $D(\omega)$ given by
Eq. (\ref{Dom}), as soon  as 
a reasonable $\Gamma'(\omega)$ is chosen. 

- The shape of the $q$-dependent part of the VH spectrum has been
discussed in detail in \cite{Drey2}. It was shown that 
Eqs. (\ref{PVH}) and (\ref{PT}) allowed to adequately describe the existence
of a Rytov dip \cite{Starunov1967} 
in a normal molecular liquid, this  dip being a 
very narrow central peak, wave-number and scattering-angle dependent,
which is subtracted from the much broader central mode. This dip
appears in the high-temperature regime when, for all the
frequencies of the central mode, $\omega \tau \ll 1$. The $\omega \tau \gg 1$
regime, which is characterised by the appearance of the Brillouin 
spectrum of a transverse propagative mode, is also well described by these 
equations, provided reasonably decreasing functions are also taken for the 
three remaining memory functions. In particular, the transverse sound 
velocity, characterised by the plateau value of $\omega \eta_T(\omega)$ 
at frequencies $1 \ll \omega \tau \ll \omega_0 \tau$ is decreased by the
coupling of the molecular orientation to the transverse phonon 
through  $\mu(\omega)^2$. 
Finally, in view of the form of $\eta_L(\omega)$, see Eq. (\ref{defetaT}),
the same is true for the sound velocity of the longitudinal phonons. 

In this Section, we shall compare the results which can be obtained 
through Eqs. (\ref{momentumcurrent}) and (\ref{osci}) 
with those resulting from the two 
other papers (or series of papers) already mentioned which make use, 
in different ways, 
of a Mori-Zwanzig 
technique to describe the liquid dynamics. 

\subsection{The Andersen-Pecora approach}
The Anderson and Pecora approach \cite{Andersen1971} 
was only used to study the VH spectrum of a molecular liquid 
at high temperature \footnote{We shall not discuss here the papers
of Keyes and Kivelson [J. Chem. Phys. {\bf 54}, 1786 (1971), {\it ibid} 
{\bf 56}, 1057 (1972)] which are, basically, along the same line.}. 
Indeed, their work was devoted to the 
explanation of the Rytov dip \cite{Starunov1967}; and their analysis 
made use of a dielectric fluctuation model 
identical to the one of the present paper.

In the work of 
Andersen and Pecora,  the mass density, the mass current and a 
second-rank tensor proportional to $Q_{ij}$ were  the sole 'slow variables'
of the theory, within the usual Zwanzig-Mori distinction between 'slow' and
'fast' variables. In other words, 
they implicitly assumed that $\dot{Q}_{ij}$ had a much faster 
dynamics than $Q_{ij}$, so that the former could be treated on the
same footing as the other fast variables. 
Furthermore, they performed a Markov approximation
on all the retarded interactions that needed to be taken into account, 
which is equivalent to taking the $\omega \tau \ll 1$ limit of the 
corresponding kernels. 

A summary of the result of their theory, within this Markov approximation,
can be found 
in the book of Berne and Pecora \cite{Berne76}. The corresponding
equations read, with notations adapted to the present paper:
\begin{eqnarray}\label{Andersen1}
\pi_{ij}(\vek{q},t) = - \Gamma_{11}'' \tau_{ij}(\vek{q},t) - 
i \Gamma_{12}' Q_{ij}(\vek{q},t ) \, ,
\end{eqnarray}
\begin{eqnarray}\label{Andersen2}
\dot{Q}_{ij}(\vek{q},t) = - i \Gamma_{21}' \tau_{ij}(\vek{q},t) 
- \Gamma_{22} Q_{ij}(\vek{q},t) \, .
\end{eqnarray}
Here, the kinetic coefficients $\Gamma_{11}'', \Gamma_{22}$ are 
real quantities, whereas $\Gamma_{12}', \Gamma_{21}'$ are purely imaginary, 
and are  related by Onsager's principle.  
Equation (\ref{Andersen1}) makes it clear that $\pi_{ij}(\vek{q},t)$ 
depends, here, linearly on $Q_{ij}(\vek{q},t)$, and not 
on its time derivative, as is the case in Eq. (\ref{trans}), while Eq. 
(\ref{Andersen2}) does not contain a second time derivative
of $Q_{ij}(\vek{q},t)$, contrary to Eq. (\ref{osci}).

The form of the Zwanzig-Mori technique, see Eq. (\ref{opid}), used in the
present paper allows to derive precise  expressions
for
 the three relaxation kernels associated with  the variables of the problem 
(time dependent generalisations of $\Gamma_{11}'', \Gamma_{12}'$ and 
$\Gamma_{22}$), 
in terms of a reduced time evolution operator. Calculating from the
corresponding equations of motion the 
VH spectrum and comparing the results with Eqs. (\ref{I_D}, \ref{I_T}), 
one can express the three Andersen-Pecora kernels as functions 
of $\eta_s(\omega), \mu(\omega), \Gamma'(\omega), \Lambda'$ and $\omega_0$. 
One can thus study their $\omega \tau \ll 1$ and $\omega \tau \gg 1$ 
regimes. This study will show that
the $\omega \tau \ll 1$ limit gives reasonable results, which are, 
as expected, in line 
with the Andersen-Pecora Markov approximation. Conversely, the 
$\omega \tau \gg 1$ limit yields a complicated behaviour for the
same three kernels which cannot be easily modeled. This will 
make the Andersen-Pecora method inappropriate  for the study
of a molecular supercooled liquid, as we shall now see.

Indeed, using the same technique as in Section II, one can derive 
the equations of motion related to $\Pi_{ij}(\vek{q},t)$ and 
$Q_{ij}(\vek{q},t)$ when one restricts
the variables to the Anderson-Pecora set. This means that, e.g. 
the projection operator $P$ of \gl{project} has been replaced by:
\begin{eqnarray}\label{Andersen3}
\hat{P} &= & | Q_{kl}(\vek{q})) \frac{1}{2S^2} (Q_{kl}(\vek{q}) | 
+ | \rho(\vek{q})) \frac{c^2}{m^2 v^2} (\rho(\vek{q}) | \nonumber \\
& & 
+ | J_k(\vek{q}) ) \frac{1}{m^2 v^2}(J_k(\vek{q}) | \, .
\end{eqnarray}
$\hat{P}$ leads to the new orthogonal projector $\hat{Q} = 1- \hat{P}$,
in terms of which a new reduced time evolution operator $\hat{R}(t)$ can 
be defined. 
One then finds that the equation for $p(\vek{q},t)$, Eq. (\ref{scalar}), 
is not 
modified, except for the change of $R'(t)$ into $\hat{R}'(t)$, which 
formally changes 
$\eta_b(t)$ into $\hat{\eta}_b(t)$. Evaluating the corresponding 
Green-Kubo relation reveals that $\hat{\eta}_b(t) = \eta_b(t)$. Since 
the pressure fluctuations are irrelevant for the VH spectrum discussed
in \cite{Andersen1971}, we do not discuss further those aspects. 
Conversely,
the equation for $\pi_{ij}(\vek{q},t)$ now turns out to be:
\begin{eqnarray}\label{Andersen5}
\pi_{ij}(\vek{q},t) &= & - \int_0^t \hat{\eta}_s(t-s) \tau_{ij}(\vek{q},s) ds
\nonumber \\
& & 
+ \int_0^t \lambda(t-s) Q_{ij}(\vek{q},s) ds + noise \, ,
\end{eqnarray}
$\hat{\eta}_s(t)$ being the complete 
analog of $\eta_s(t)$. Equation (\ref{Andersen5}) 
contains a linear term in $Q_{ij}(\vek{q},t)$, as in
Eq. (\ref{Andersen1}), and not in $\dot{Q}_{ij}(\vek{q},t)$, as
was case for Eq. (\ref{trans}) while
the corresponding retarded interaction is expressed
by:
\begin{eqnarray}
( \dot{Q}_{ij} | \hat{R}'(t) | \pi_{kl} ) \frac{1}{S^2} = - \lambda(t) 
\Delta_{ij,kl} \, ,
\end{eqnarray}
where $\dot{Q}_{ij}$ and $\pi_{kl}$ are respectively odd and even
with respect to time inversion;  thus $\lambda(t)$ is an odd function of
$t$, contrary to all the memory functions considered up to now in the
present paper. The most important change arises, nevertheless,
 from the fact that 
the equation of motion for $Q_{ij}(\vek{q},t)$ has to be derived from: 
\begin{equation}
\partial_t Q_{ij}(\vek{q},t) = \dot{Q}_{ij}(\vek{q},t) \, ,
\end{equation}
an equation which replaces Eq. (\ref{ddotQ}). A calculation similar in every
respect to the one performed below Eq. (\ref{momentumcurrent}) yields:
\begin{eqnarray}\label{Andersen7}
\dot{Q}_{ij}(\vek{q},t) &= & -\frac{\Lambda'}{\omega_0^2} \int_0^t 
\lambda(t-s) \tau_{ij}(\vek{q},s) ds \nonumber \\
& & - \int_0^t M(t-s) Q_{ij}(\vek{q},s) ds +noise \, , 
\end{eqnarray}
with:
\begin{eqnarray}
( \dot{Q}_{ij} | \hat{R}'(t) | \dot{Q}_{kl} )\frac{1}{S^2}
 = M(t) \Delta_{ij,kl} \, .
\end{eqnarray}
Equations (\ref{Andersen5}) and (\ref{Andersen7}) are, obviously, the 
non-Markovian form of Eqs. (\ref{Andersen1}) and (\ref{Andersen2}). Ignoring
temperature fluctuations, the previous relations are exact, and  
allow us to  relate the memory kernels 
$M(\omega), \hat{\eta}_s(\omega), \lambda(\omega)$ to the ones already 
used in
this paper by deriving from Eqs. (\ref{Andersen5},\ref{Andersen7}), 
through the same methods as used in \cite{Drey2}, the expression of the
VH spectrum and comparing it with Eq. (48) of Part I.  
 For $M(\omega)$, this can be done by simply computing 
the correlation function of $Q_{\perp\perp'}$ 
which is  responsible for the pure back-scattering 
spectrum (see Part I for a definition of the geometry used).
Solving Eq. (\ref{Andersen7}) in this simple case yields:
\begin{eqnarray}
LT[ (Q_{\perp \perp'}(\vek{q},t) | Q_{\perp \perp'}(\vek{q}) ) ](\omega) =
\frac{S^2}{\omega- M(\omega)} \, .
\end{eqnarray}
Comparison with Eq. (34) of Part I, with the help of Eq. (\ref{S}), 
leads to the relation 
 between $\Gamma'(\omega)$ 
and $M(\omega)$: 
\begin{eqnarray}\label{MGam}
M(\omega) = \frac{\omega_0^2}{\omega - \Gamma'(\omega)} \, .
\end{eqnarray}

In order to gain some insight into the Markov approximation, a priori 
valid at high temperatures, let
us discuss the properties of $M(\omega)$ upon cooling the system. To simplify
the discussion, we consider a Maxwell model for $\Gamma'(\omega)$:
\begin{eqnarray}\label{model}
\Gamma'(\omega) = i \gamma + \frac{i \Gamma_0^2 \tau}{1+ i \omega \tau} \, ,
\end{eqnarray}
which mimics its frequency dependence for frequencies much lower than 
typical liquid frequencies, i.e. for
 $\omega \ll \omega_0$. Then, all the fast processes
are hidden in a weakly temperature-dependent background, $i \gamma$, whereas
the temperature-sensitive structural relaxation is modeled by a decreasing
exponential in the time domain, corresponding to a temperature insensitive
amplitude, $\Gamma_0^2$, and a relaxation time, $\tau$, 
that increases by orders
of magnitude upon supercooling the liquid. 

At high temperature, i.e. for $\omega \tau \ll 1$, under the 
possible conditions $\omega, \gamma, \omega_0 \ll \Gamma_0^2 \tau$, 
which simply 
imply that even at high 
temperatures $\Gamma_0^2 \tau$ is substantially larger 
than the width of the central peak, Eq. (\ref{I_D}), one obtains:
\begin{equation}\label{Mhigh}
M(\omega) \simeq i \frac{\omega_0^2}{\Gamma_0^2 \tau} \, .
\end{equation}
This is the Andersen-Pecora result in its Markov approximation
$M(\omega) = i \Gamma_{22}$ with $\Gamma_{22} \sim \tau^{-1}$. 
Conversely, at low temperature, and under the same 
conditions, $M(\omega)$ will be approximated by:
\begin{equation}
M(\omega) \simeq \frac{- \omega_0^2 \omega}{\Gamma_0^2 + i \omega \gamma}
\simeq - \frac{\omega_0^2 \omega}{\Gamma_0^2} \, ;
\end{equation}
at low frequencies, $M(\omega)$ is a real quantity proportional to $\omega$
 and independent of the relaxation time. 
The vanishing of the imaginary part of $M(\omega)$ 
at low frequencies implies that 
the area of $M(t)$ cancels at low temperatures. 
Whereas the shape of 
$M(t)$ is clearly model dependent, 
the cancellation of areas of $M(t)$ is a general feature 
 of supercooled liquids. Contrary to 
$\Gamma'(t)$, there is no step process in $M(t)$ with a diverging time scale 
upon cooling. 
It is thus fruitless to try to model the temporal
evolution of $M(t)$ since the common features of an increasing structural 
relaxation time are masked in this approach.  

We can similarly evaluate 
 the Andersen-Pecora 
memory kernels $\hat{\eta}_s(\omega)$, and  $\lambda(\omega)$ 
by solving 
the dynamics for the variable $Q_{\perp\parallel}(\vek{q},t)$ 
that also contributes to the VH scattering. Using the methods of 
\cite{Drey2}, 
one obtains:
\begin{eqnarray}\label{Andersen17}
& & 
LT[ (Q_{\perp \parallel}(\vek{q},t) | Q_{\perp \parallel}(\vek{q}) ) ](\omega)
 =  
\frac{S^2}{\omega- M(\omega)}  \nonumber \\
& & -  
\left[  \frac{\Lambda'}{mn \omega_0^2}  
\frac{\lambda(\omega)}{\omega- M(\omega)} \right]^2 
\frac{q^2 m^2 v^2}{ \omega -q^2 \eta_T(\omega)/mn} 
\, ,
\end{eqnarray}
with the transverse viscosity given by:
\begin{eqnarray}
\eta_T(\omega) = \hat{\eta}_s(\omega) - \frac{\Lambda'}{\omega_0^2} 
\frac{\lambda(\omega)^2}{\omega - M(\omega)} \, .
\end{eqnarray}
A comparison between the second term of the r.h.s of
Eq. (\ref{Andersen17}) and Eq. (\ref{I_T}) yields the relation between the 
two sets of memory kernels:
\begin{eqnarray}\label{lam}
 \lambda(\omega) = - i \frac{\omega_0^2 \mu(\omega)}{\omega- \Gamma'(\omega)} 
= -i \mu(\omega) M(\omega)
\, , 
\end{eqnarray}
\begin{eqnarray}\label{hateta}
\hat{\eta}_s(\omega) &= &  \eta_s(\omega) + \frac{\Lambda' \mu(\omega)^2}{\omega- 
\Gamma'(\omega)} \nonumber \\
& = &  \eta_s(\omega) + \frac{\Lambda'}{\omega_0^2} \mu(\omega)^2 M(\omega) 
\, .
\end{eqnarray} 
The high-temperature limit of $M(\omega)$, Eq. (\ref{Mhigh}), yields, with 
a Debye model for $\mu(\omega)$ and ${\eta}_s(\omega)$, 
the Markov limits obtained in 
\cite{Andersen1971}: $\lambda(\omega)$
becomes an imaginary number independent of $\tau$, while 
$\hat{\eta}_s(\omega \to 0)$ is the sum of two terms, both imaginary
and proportional to the relaxation time. 
One originates from $\eta_s(\omega \to 0)$, while the second, negative,
is the $\omega \to 0$ limit of $\Lambda' \mu(\omega)^2 M(\omega)/\omega_0^2$;
this explains why, in the Andersen-Pecora approach, the viscosity, 
$-i \eta_s(\omega\to 0)$, is the sum of two positive terms. 

Conversely, the difficulty of an a priori modeling of $M(\omega)$ 
transfers to the two other memory kernels, $\hat{\eta}_s(\omega)$ and
$\lambda(\omega)$. 
This explains why a na{\"\i}ve modeling by simple, 
Debye-like, relaxation functions, consistent both with 
the high-temperature Markov results and the different 
time reversal symmetries of the
kernels, is unable to yield correct physical results in the
supercooled regime. Appendix D shows, indeed, that the low-temperature
limit of such an attempt leads to the existence of 
transverse propagative modes coupled to molecular orientation motions,
but this coupling increases the sound velocity instead of decreasing
it. 

Summarizing this part, we have shown that a Mori-Zwanzig procedure as
applied by Andersen and Pecora \cite{Andersen1971} 
allows to derive constitutive
equations through which the light scattering problem can be 
properly formulated as long
the frequency dependence is kept on a formal level. 
Conversely, when one expresses these memory kernels in terms of those
obtained in Section II, one discovers that their 
modeling as  time-dependent memory
kernels is extremely difficult. This problem can be circumvented, 
if  one models
them 
directly, as inspired by Eqs. (\ref{MGam}), 
(\ref{lam}), (\ref{hateta}), but this  procedure is equivalent to considering 
$\Gamma'(\omega), \mu(\omega)$, and $\eta_s(\omega)$ as fundamental
quantities.

\subsection{Comparison with  the general expressions for light scattering}
The expressions for the VV and VH intensities
obtained in Part I and discussed again in Section IV, Eqs. (\ref{PVV}) and 
(\ref{PVH}), have been obtained under the physical assumption that the
fluctuations of the local dielectric tensor, 
$\delta \epsilon_{ij}(\vek{q},t)$,
could be expressed through \gl{scattering_model}, i.e. 
depend, in first order, only on the density and the 
orientational fluctuations.
Conversely, the expressions obtained in \cite{Franosch2001} did 
not make use of a specific form for $\delta \epsilon_{ij}(\vek{q},t)$. 
We shall show, in this last part of Section V, that those two expressions are,
indeed, a specialization of the
 general results obtained in \cite{Franosch2001}. We also discuss the 
respective merits of these two complementary approaches. 

The  basic idea of \cite{Franosch2001} was  to express the
 finite wave-number fluctuations of the 
dielectric tensor in terms of the long-wavelength limit of the two 
special linear combinations:
\begin{eqnarray}
s_{00}(\vek{q}) & = & [\epsilon_{xx}(\vek{q}) +
 \epsilon_{yy}(\vek{q}) + \epsilon_{zz}(\vek{q}) ]/3 \, , \nonumber \\
t_{20}(\vek{q})& = & [2 \epsilon_{zz}(\vek{q}) - \epsilon_{xx}(\vek{q}) 
 - \epsilon_{yy}(\vek{q}) ]/\sqrt{12} \, .
\end{eqnarray} 
Within the present light-scattering model, 
 these two quantities reduce to the contributions 
of the density and the orientation fluctuations, respectively:
\begin{subequations}
\begin{eqnarray}
\label{s&t}
s_{00}(\vek{q}) &= &  a \rho(\vek{q}) \, , 
\end{eqnarray}
\begin{eqnarray}\label{t&t}
t_{20}(\vek{q})  &= &   
b Q_0(\vek{q}) \, .
\end{eqnarray} 
\end{subequations} 
In \cite{Franosch2001}, 
the VV spectrum was expressed, using   notations that 
 have been adapted to the current paper, as: 
\begin{eqnarray}
\label{FVV}
& & I_{VV}(\vek{q},\omega) = Im \Bigg\{
{\cal S}(\omega) + \frac{4}{3} {\cal T}(\omega) +\frac{m^2 v^2}{c^2 \omega}
\left( \frac{\partial s_{00}}{\partial \rho} \right)_T^2 
\nonumber \\
& & + \left[ \left. \frac{\partial s_{00}}{\partial \rho} \right)_T 
- \omega a_{VV}(\omega) \right]^2 
\left[ LT[ (\rho(\vek{q},t)| \rho(\vek{q}))](\omega) 
-\frac{m^2 v^2}{c^2 \omega} \right]
\nonumber \\
& & + \xi(\omega)^2 LT[(\Theta(\vek{q},t) | \Theta(\vek{q}))](\omega) 
\nonumber \\
& & + 2 \xi(\omega) 
\left[ \left( \frac{\partial s_{00}}{\partial \rho} \right)_T 
- \omega a_{VV}(\omega) \right] 
LT[ (\rho(\vek{q},t)| \Theta(\vek{q}))](\omega) \Bigg\} \, .
\end{eqnarray}
Similarly, the
VH intensity was expressed as:
\begin{eqnarray}\label{omIVH}
& & I_{VH}(\vek{q},\omega) = Im \Bigg\{ 
{\cal T}(\omega)   \nonumber \\
& &  
+ \frac{q^2}{\omega} \cos^2 \frac{\theta}{2} 
[\omega a_{VH}(\omega)]^2 m^2 v^2 P_T(q,\omega)
\Bigg\} \, .
\end{eqnarray} 
The transverse phonon propagator is 
given directly in terms of the transverse viscosity, $\eta_T(\omega)$, 
as in  Eq. (\ref{PT}). 
As already alluded in the Introduction, 
the price to be paid for these two general results was the introduction
of ten  
frequency-dependent quantities, namely, the scalar background spectrum 
${\cal S}(\omega)$, the tensor background spectrum 
${\cal T}(\omega)$, the two Pockels' coupling functions,
 $a_{VV}(\omega)$ 
for polarised
and 
$a_{VH}(\omega)$ for depolarised scattering, and the temperature coupling 
$\xi(\omega)$, while the three  hydrodynamic correlation functions 
related to the density, $\rho(\vek{q})$, and 
kinetic temperature, $\Theta(\vek{q})$, were
 expressed in terms of the transverse viscosity, $\eta_T(\omega)$,
the longitudinal viscosity, $\eta_L(\omega)$, the heat conductivity,
 $\lambda(\omega)$, the dynamic specific heat, $c_V(\omega)$,
 and  the tension coefficient, $\beta(\omega)$, respectively. 
The singular hydrodynamic behavior  manifested itself explicitly in the 
three correlation functions just mentioned.

Let us relate those quantities to the one derived in the present paper and 
demonstrate numerous simplifications that occur in the 
density-and-orientational-decay-channels-only model. 
First, the temperature coupling is given by:
\begin{equation}\label{xi2}
\xi(\omega) = \xi
\frac{c_V(\omega)}{ c_V} - \frac{c_V(\omega)}{c_V^0} \omega \frac{ 
LT[(\tilde{Q}
s_{00}(t) | \tilde{Q}e^P)](\omega)}{k_B  T^2} \, ,
\end{equation}
where $\xi$ is a linear function of the energy fluctuations in the
$q\to 0$ limit and  
$\tilde{Q} = 1 -\tilde{P}$ is a projection operator 
orthogonal to the standard 
Kadanoff-Martin projector, $\tilde{P}$. The latter
  projects, in \cite{Franosch2001}, on  five
variables, $\rho(\vek{q})$ and  $\vek{J}(\vek{q})$, as in the present 
paper,  and on the temperature
fluctuations, $T(\vek{q})$, not introduced here, and 
proportional to the energy fluctuations, $e(\vek{q})$.
\begin{eqnarray}
\tilde{P} & = & | \rho(\vek{q}) ) \frac{c^2}{m^2 v^2} (\rho(\vek{q}) | 
+ |J_k(\vek{q}) ) \frac{1}{m^2 v^2} (J_k(\vek{q}) | \nonumber \\
& &
+ | T(\vek{q}) ) \frac{c_V}{k_B T^2} (T(\vek{q}) |
+ {\cal O}(q^2) \,  .
\end{eqnarray}
Because energy fluctuations are not considered 
in the present approach, $\xi \equiv 0$. Also, $s_{00}(\vek{q})$ is 
proportional to $\rho(\vek{q})$, \gl{s&t}, so that $\tilde{Q} s_{00} = 0, 
\xi(\omega) \equiv 0$.

Second, 
in \cite{Franosch2001}, the scalar and the tensor background spectra
are defined as:
\begin{subequations}
\begin{eqnarray}
 {\cal S}(\omega) & = &  -\frac{ k_B T^2 \xi(\omega)^2 }{\omega c_V(\omega)} + \frac{ k_B T^2}{\omega c_V}
\xi^2 \nonumber \\
& & 
 + LT[( s_{00}(t) | \tilde{Q} s_{00}) ](\omega) \, , 
\end{eqnarray}
\begin{eqnarray}
{\cal T}(\omega) & = &  LT[(t_{20}(t) |  t_{20})](\omega) \, .
\end{eqnarray}
\end{subequations}
The preceeding results imply:
\begin{subequations}
\begin{eqnarray}
{\cal S}(\omega) & \equiv &  0 \, , 
\end{eqnarray}
\begin{eqnarray}\label{tensorspectrum}
{\cal T}(\omega) &= &  b^2 LT[ (Q_0(t) | Q_0 )](\omega) \, .
\end{eqnarray}
\end{subequations}
Third, from \gl{s&t},  
$(\partial s_{00}/\partial \rho)_T 
= a$ while  
the dynamic Pockels' coupling functions are given, in \cite{Franosch2001},
as:
\begin{subequations}
\begin{eqnarray}\label{aVH}
a_{VH}(\omega) &= & LT[ (\pi_{20}(t) | t_{20})](\omega)/m^2 v^2
 \, , 
\end{eqnarray}
\begin{eqnarray}
a_{VV}(\omega) & = & 
\frac{2}{3} a_{VH}(\omega) 
+ \frac{\beta(\omega) T \xi(\omega)}{\omega c_V(\omega)} 
- \frac{\beta T \xi}{\omega c_V} \nonumber \\
& &
- LT[(p(t) | \tilde{Q} s_{00})](\omega) /m^2 v^2 \, .
\end{eqnarray}
As $\xi(\omega), \xi$ and $\tilde{Q} s_{00}$ are all equal to 
zero, one
obtains:
\begin{equation}
a_{VV}(\omega) = \frac{2}{3} a_{VH}(\omega) \, .
\end{equation}
\end{subequations}
Let us look at the results for the VV light scattering intensities. 
One observes that terms involving dynamic correlation functions of
the kinetic temperature evaluate to zero. One is thus left with
the density correlation functions which,
if energy fluctuations are ignored, reads
in agreement with Eq. (29a) of Part I:
\begin{eqnarray}\label{LTrho}
& & LT[(\rho(\vek{q},t) | \rho(\vek{q}))](\omega) = \nonumber \\
& & 
\frac{m^2 v^2}{\omega} \left[ \frac{1}{c^2} + 
\frac{q^2}{\omega^2 -q^2 [c^2 
+ \omega \eta_L(\omega)/mn]} \right] \, .
\end{eqnarray}
Equation
(\ref{LTrho}) 
allows to group the terms 
proportional to $m^2 v^2$ and one ends up with:
\begin{eqnarray}\label{omIVV}
& & I_{VV}(\vek{q},\omega) 
=   Im \Bigg\{ 
\frac{4}{3} {\cal T}(\omega) + \frac{a^2 m^2 v^2}{c^2 \omega} \nonumber \\
& & 
+  \left[a - \frac{2}{3} \omega a_{VH}(\omega) \right]^2
\frac{m^2 v^2 q^2/\omega }{\omega^2 - q^2 c^2 - q^2 \omega \eta_L(\omega)/mn}
\Bigg\} \, ,
\end{eqnarray}
${\cal T}(\omega)$ and $a_{VH}(\omega)$ being defined through 
Eqs. (\ref{tensorspectrum}) and (\ref{aVH}),
respectively. 
Furthermore in agreement with \gl{Q0Q0} and Eq. (34) of Part I, 
${\cal T}(\omega)$ can be written as
\begin{equation}\label{omT}
{\cal T}(\omega) = b^2 \frac{S^2}{\omega} \left( 1 - 
\frac{\omega_0^2}{D(\omega)} \right) \, ,
\end{equation}
which is here a simple definition of $\omega_0^2/D(\omega)$. 
In the same manner, 
from Eqs. (\ref{aVH},\ref{t&t},\ref{romega},\ref{lambda}), 
one obtains:
\begin{equation}\label{omaVH}
- \omega a_{VH}(\omega) = \frac{b \Lambda'}{mn} r(\omega) \, ,
\end{equation}
where, similarly to the case of $\omega_0^2/D(\omega), r(\omega)$ 
is simply defined through the Laplace transform of $(\pi_{20}(t) |t_{20} )$,
\gl{aVH}. One sees that \gl{omIVV}
has been cast into a form identical to \gl{PVV}, while 
a similar identification holds between Eq. (\ref{omIVH}) and 
Eq. (\ref{PVH}).

The reduction of Eqs. (\ref{FVV}) and (\ref{omIVH})
 to Eqs. (\ref{PVV}) and (\ref{PVH})
 shows the comparative interests
of the approach of \cite{Franosch2001} and of the present one. 
The method of \cite{Franosch2001}
does not depend on the system under study and allows for temperature 
(or energy) fluctuations: as soon as the scattering  
model, Eq. (\ref{scattering_model}),  
 is introduced, and the energy fluctuations are neglected, the equations 
of \cite{Franosch2001} reduce to those of the present model,
depending on four functions ${\cal T}(\omega), a_{VH}(\omega), 
\eta_L(\omega)$ and 
$\eta_T(\omega)$, which are undetermined in this framework. Conversely,
the more restricted approach developed in the present series of papers
gives precise definitions of these four quantities in terms 
of more fundamental memory kernels $\Gamma'(\omega), \mu(\omega), 
\eta_b(\omega)$ and $\eta_s(\omega)$, and also gives the relationships 
through which the four first functions are related to the 
second ones via the constants $\Lambda'$ and $\omega_0$ for 
which definitions can be given. Yet the restricted approach has its own 
price to be paid: one has to start the whole work again if additional 
variables need to be introduced into the model.

\section{Summary and final remarks}
In a liquid formed of rigid molecules, the dynamics of the system has to take
into account both the motion of the molecular centers of mass
and the orientational motion of the molecules. In the
long-wavelength limit, the first one gives rise to the hydrodynamic
modes, to which the orientational motions are partly coupled, while this 
orientational dynamics also gives rise to motions that are 
wave-vector independent in the same limit. \cite{Drey1,Drey2} and 
\cite{Dreyfus2002a} proposed a phenomenological set of 
equations to describe this coupled dynamics in the 
case of linear molecules, and a phenomenological expression
for the local fluctuation of the dielectric tensor: this fluctuation
was expressed in terms of the density and orientational variables entering
the dynamical equations.

The original objective of the present paper was twofold -- one was to 
provide a complete, microscopic, derivation of these dynamical 
equations; the second was to compare the expression 
for the light scattering
intensities resulting from these equations with those obtained with  two 
other approaches \cite{Andersen1971,Franosch2001}.

Both goals have been achieved. On the one hand, the use of 
a Zwanzig-Mori formalism has allowed to completely derive 
these dynamical equations; in the course of this derivation,
we have obtained the microscopic
expressions of the two parameters and of the four memory functions 
entering those equations. On the other hand, the comparison with the two other 
Zwanzig-Mori approaches has also brought important results. 
One of them is related to the choice of Andersen and Pecora 
\cite{Andersen1971} of not including $\dot{Q}_{ij}(\vek{q})$ 
in their set of variables. This choice, which is sufficient at high 
temperature, when the Markov approximation can be made on the
corresponding memory kernels, 
turns out to be inappropriate for the study of supercooled
liquids: at low temperatures $\dot{Q}_{ij}(\vek{q},t)$ is as
'slow' a variable as $Q_{ij}(\vek{q},t)$.
The most important consequence of the absence 
of $\dot{Q}_{ij}(\vek{q},t)$ in the set of selected variables is the
change in the equation of motion of $Q_{ij}(\vek{q},t)$: it transforms
it from a second order differential equation with a memory kernel
acting on $\dot{Q}_{ij}(\vek{q})$ into a first order differential
equation with a memory kernel acting on $Q_{ij}(\vek{q})$.
The formal neglect of $\dot{Q}_{ij}(\vek{q})$
in the set of 'slow' variables  
is possible but the corresponding memory kernels
have a non-trivial time evolution, which cannot  be predicted without 
using the 
results of the present theory. The second result is that the present 
formulation of the theory is, indeed, a reduction of the 
general theory of \cite{Franosch2001} which can be derived from  
simplifications 
consistent with the phenomenological 
model of the dielectric tensor, and with the 
restricted set of variables used here.  

The present Zwanzig-Mori approach also led 
to two important byproducts. 
One is the existence of conditions, Eqs. (\ref{positive},\ref{Cauchy}), 
which have to be fulfilled by 
the Laplace transforms
of the memory functions. These 
conditions are important because they are sufficient to insure that
 all the light scattering intensities will be positive, whatever the 
frequency, within the phenomenological model of the fluctuations of the 
dielectric  tensor used here. 
A 
second byproduct is the set of Green-Kubo 
formulae we have derived in Section
IV-C: we have shown that the correlation functions of some variables, 
not experimentally accessible by light scattering techniques, 
but which may be numerically obtained from MD 
computations of models of these molecular (supercooled) liquids, give access
to definite combinations of the Laplace transforms of these memory
functions. This is a possible way of obtaining an information on them. 

Some results of the present paper provide a direct help to the 
experimentalists, when analysing the light scattering spectra
of molecular supercooled liquids formed of rigid linear molecules, or 
of molecular liquids for which such  an approximations is reasonable. 
One of them is the already mentioned necessary conditions on the
memory functions. A second is that these functions exhibit 
the characteristic features of structural relaxation, e.g. 
rapidly increasing   relaxation times upon lowering the
temperature. 
Yet, the functional form of these memory kernels remains undetermined 
within this framework, except for the conventional analytic properties. 
Usually it is not possible to directly extract the frequency dependence of 
of the memory kernels from light-scattering experiments. Rather one has
to rely on empirical functions and adjust a small number of
 parameters to obtain
a reasonable description of experimental data. As a further step, one can 
supplement these empirical functions 
with features inspired from theoretical considerations, e.g.
the fast $\beta$-process as discussed in the context of mode-coupling theory
 \cite{Goetze1999}.

Time resolved optical spectroscopy of the same molecular 
liquids has recently
developed into an important tool; this is particularly the case for 
the impulsive stimulated thermal scattering technique (ISTS) mentioned
in the Introduction \cite{Yang95,Paolucci2000,Torre2001}. 
The most important 
part of the new information 
obtained from these measurements
is derived from the coupling of the heat diffusion process with the
stimulated hydrodynamics mode. We have not incorporated, in the 
microscopic derivation of the dynamical equations, a local 
temperature as a pertinent
variable, contrary to what has been done in \cite{Franosch2001}. 
In order to properly exploit the information contained in these ISTS 
experiments, the whole procedure developped in the present paper has to 
be repeated with the inclusion of the variable(s) describing the local 
temperature of the supercooled liquid. It has to be found if this 
generalisation will require more memory functions than could be anticipated
from a phenomenological extension of the full set of 
Navier-Stokes equations to the case of a supercooled (memory 
function aspect) molecular (inclusion of the rotation-translation coupling 
and of the molecular orientation dynamics) liquid \cite{Dreyfusprep}.

\begin{acknowledgments}

We wish to thank  H.~Z. Cummins and W. G{\"o}tze for their useful
comments and suggestions. 
\end{acknowledgments}

\appendix
\section{Static averages} 
The Hamilton  function of identical, interacting,  symmetric tops reads:
\begin{equation}\label{A1}
H = \sum_\alpha T_\alpha + V(\{\vek{R}_\alpha,\phi_\alpha,\Theta_\alpha \}) 
\, ,
\end{equation}
where the kinetic energy of the $\alpha$-th  molecule is given by:
\begin{equation}\label{A2}
T_\alpha = \frac{\vek{P}_\alpha^2}{2m} + 
\frac{(p_{\alpha\phi} - p_{\alpha\psi} 
\cos \Theta_\alpha)^2}{2 I \sin^2 \Theta_\alpha} 
+ \frac{p_{\alpha\Theta}^2}{2 I} + \frac{p_{\alpha \psi}^2}{2 I'} \, .
\end{equation}
Here $\vek{R}_\alpha, \phi_\alpha, \Theta_\alpha, \psi_\alpha$ 
denote the center-of mass position and the Euler angles of the molecule
following the definition of \cite{Gray}, and
$\vek{P}_\alpha, p_{\alpha\phi}, p_{\alpha\Theta}, p_{\alpha\psi}$ 
the corresponding canonical momenta. The moments of 
inertia are denoted by $I,I'$ for rotation perpendicular
 to and around the molecule's axis of symmetry. The
potential energy of the interacting molecules is denoted by $V$. 
Note that, due to the symmetry,
 the interaction does not depend on the Euler angles $\psi_\alpha$. 

The orientational current, $\dot{Q}_{ij}(\vek{q}) = i {\cal L}  
Q_{ij}(\vek{q}) = \{ H , Q_{ij}(\vek{q}) \}$,
 then splits naturally into 
two parts:
\begin{subequations}
\begin{eqnarray}\label{dotQexplicit}
\dot{Q}_{ij}(\vek{q}) &= &  N^{-1/2}  
\sum_{ \alpha =1}^N \frac{i \vek{q} \cdot \vek{P}_\alpha}{m} 
\left( \hat{u}_{\alpha i} \hat{u}_{\alpha j} - \frac{1}{3} \delta_{ij} \right) 
e^{ i \vek{q} \cdot \vek{R}_\alpha } \nonumber \\
& & +N^{-1/2}  
\sum_{ \alpha =1}^N e^{ i \vek{q} 
\cdot \vek{R}_\alpha } i {\cal L}  \hat{u}_{\alpha i} \hat{u}_{\alpha j}  
 \, ,
\end{eqnarray}
where the first term corresponds to the translational motion of the 
center of mass and  the second term  describes molecular 
 reorientations. 

By definition:
\begin{eqnarray} 
( A(\vek{q}) | B(\vek{q}) ) & =& 
\int d \Gamma e^{-H/k_B T}  \delta A(\vek{q})^*   \delta B(\vek{q}) \, , 
\end{eqnarray}
where $d\Gamma = \prod_\alpha d\vek{R}_\alpha d\phi_\alpha d\Theta_\alpha 
d\psi_\alpha d\vek{P}_\alpha d p_{\alpha \phi} d p_{\alpha \Theta}
d p_{\alpha \psi}$ denotes the canonical phase space volume element.
Let us compute $(\dot{Q}_{ij}(\vek{q}) | J_k(\vek{q}))$. In this 
thermal average appear two types of integrals (see \gl{dotQexplicit}): 
one involving $(\vek{q} \cdot \vek{P}_\alpha) P_{\beta k}$, and the second
$(i {\cal L} \hat{u}_{\alpha i } \hat{u}_{\beta j}) P_{\beta k}$. As
\begin{equation}
i {\cal L} \hat{u}_{\alpha i } \hat{u}_{\beta j} = 
\{ \sum_\gamma T_\gamma , \hat{u}_{\alpha i } \hat{u}_{\beta j} \}
\end{equation}
involves only the angular variables, the only part dependent on  the linear
momentum in the second terms reads (c.f. Eqs. \ref{A1} and \ref{A2}):
\begin{equation}
\int e^{-P_{\beta_ k}^2/(2 m k_B T)} P_{\beta k} d P_{\beta k}
= 0 \, .
\end{equation}
\end{subequations}
One is thus left with the contributions of the first term. They read: 
\begin{eqnarray}\label{A4} 
\lefteqn{( \dot{Q}_{ij}(\vek{q}) | J_k(\vek{q}) )  =  \nonumber } \\
& & N^{-1} \sum_{\alpha,\beta=1}^N \left( 
\frac{i \vek{q} \cdot \vek{P}_\alpha}{m} \left( \hat{u}_{\alpha i} 
\hat{u}_{\alpha j} - \frac{1}{3} \delta_{ij} \right) 
e^{ i \vek{q} \cdot \vek{R}_\alpha }  \Big| P_{\beta k} e^{i \vek{q} 
\cdot \vek{R}_\beta} \right) \nonumber \\
& = & - i q_k k_B T 
N^{-1} \sum_{\alpha=1}^N  \left(  \left( \hat{u}_{\alpha i} 
\hat{u}_{\alpha j} - \frac{1}{3} \delta_{ij} \right) 
e^{ i \vek{q} \cdot \vek{R}_\alpha }  \Big|  e^{i \vek{q} 
\cdot \vek{R}_\alpha} \right) \nonumber \\
& = & -i q_k k_B T 
N^{-1} \sum_{\alpha=1}^N  \langle ( \hat{u}_{\alpha i} 
\hat{u}_{\alpha j} - \frac{1}{3} \delta_{ij} )  \rangle  = 0 \, ,
\end{eqnarray}
where the last but one equality originates from averaging over the gaussian
variable, $\vek{P}_\alpha$, and the last one from the rotational symmetry 
of the problem. Equation (\ref{A4})
has been reported as \gl{cross} in the body of the present paper.

The kinetic energy expressed in terms of canonical momenta 
depends explicitly on the Euler angles, hence the evaluation of 
thermal averages is quite involved. This can be avoided by eliminating the 
canonical momenta in favour of  the angular momenta  \cite{Gray}:
\begin{eqnarray}
p_{\alpha\phi} &= &  - \tilde{J}_{\alpha x} 
\sin \Theta_\alpha \cos \psi_\alpha 
+ \tilde{J}_{\alpha y} \sin \Theta_\alpha \sin \psi_\alpha \nonumber \\
& & + \tilde{J}_{\alpha z} \cos \Theta_\alpha \, ,\nonumber \\
p_{\alpha \Theta} & = & \tilde{J}_{\alpha x} \sin \psi_\alpha 
+ \tilde{J}_{\alpha y} \cos \psi_\alpha \, ,\nonumber \\
p_{\alpha \psi} &= & \tilde{J}_{\alpha z} \, . 
\end{eqnarray}
One checks that the Jacobian is $\sin \Theta_\alpha$, while 
 the corresponding part of the kinetic 
energy reads
$ T_\alpha = (\tilde{J}_{\alpha x}^2 +\tilde{J}_{\alpha y}^2)/2 I 
+ \tilde{J}_{\alpha z}^2/2 I'$. 
Then the partition sum is given by
\begin{equation}
Z = \int \prod_\alpha d \vek{R}_\alpha d\phi_\alpha d\cos \Theta_\alpha d \psi_\alpha d \vek{P}_\alpha d \tilde{\vek{J}}_{\alpha} e^{- H/k_B T} \, .
\end{equation}
Thus, averaging over the angular momenta is just gaussian and averaging over 
the Euler angles amounts to averaging over the usual Haar measure of the 
rotation group.

In order to calculate the long-wavelength limit of  the 
auto-correlation function of the 
orientational currents, it is sufficient to calculate 
it for one of its  components, say $\dot{Q}_{zz}(\vek{q})$. Since 
$\hat{u}_{\alpha z} = \cos \Theta_\alpha$ and $\dot{\Theta}_{\alpha} 
= \{ H , \Theta_\alpha \} = p_{\alpha \Theta }/I$, one finds, for $q\to 0$:
\begin{eqnarray}
\dot{Q}_{zz}(\vek{q}) & =  & 
- \frac{2}{I} N^{-1/2} \sum_{\alpha=1}^N [ \tilde{J}_{\alpha x} 
\sin \psi_\alpha 
\nonumber \\ & & 
+ \tilde{J}_{\alpha y} \cos \psi_\alpha ]
\cos \Theta_{\alpha} \sin \Theta_\alpha \, .
\end{eqnarray}  

Then the long-wavelength correlation function of the orientational 
current can be evaluated:
\begin{eqnarray}\label{evOm}
\lefteqn{ ( \dot{Q}_{zz}(\vek{q}=0) | \dot{Q}_{zz}(\vek{q}=0) ) 
  = } \nonumber \\
& &  
\frac{4}{I^2} N^{-1} \sum_\alpha
\langle 
\left[ \tilde{J}_{\alpha x} 
\sin \psi_\alpha + \tilde{J}_{\alpha y} \cos \psi_\alpha \right]^2  
\cos^2 \Theta_\alpha \sin^2 \Theta_\alpha \rangle 
\nonumber \\
& = & 4 \frac{k_B T}{I} N^{-1} \sum_\alpha \langle \cos^2 \Theta_\alpha
 \sin^2 \Theta_\alpha \rangle = 
\frac{8 k_B T}{15 I} \, .
\end{eqnarray} 
Comparison with \gl{Omega} yields for the ideal gas libration frequency 
$\Omega^2 = 2 k_B T/5 I$.

\section{Operator identity}
The time evolution operator, $R(t) = \exp (i {\cal L}t)$, 
may be  split into two parts
$R(t) = R_P(t) + R_Q(t)$ with $R_P(t) = R(t) P, R_Q(t) = R(t) Q$. From the 
equation of motion, $\partial_t R(t) = R(t) i {\cal L}$, one finds:
\begin{equation}\label{B1}
\partial_t R_Q(t) = R_P(t) i {\cal L} Q + R_Q(t) i {\cal L} Q \, .
\end{equation}
The solution of \gl{B1} can be expressed in terms of $R_P(t)$ as:
\begin{equation}
R_Q(t) = Q e^{i {\cal L} Q t} 
+ \int_0^t R_P(s) i {\cal L} Q e^{i {\cal L} Q (t-s)}
ds \, .
\end{equation}
Furthermore, because $R'(t) = Q e^{i { \cal L} Q t}$ incorporates the 
projection operator $Q$, one easily finds, by e.g. expansion of the 
exponential, that $R'(t)$ may be written in the symmetric form: 
\begin{equation}
R'(t) 
= Q e^{i Q{\cal L} Q t} Q \, .
\end{equation}
Collecting terms, one arrives at \gl{opid}. 

\section{Properties of the memory functions}
The memory kernels of the type $(A | R'(t) | A )$ 
exhibit the same mathematical properties as auto-correlation 
functions, viz.
for complex frequencies in the lower half plane, their 
 Laplace transform is 
analytic with non-negative imaginary part. 
A non-rigorous proof can be adapted from Berne and Pecora \cite{Berne76}. 
Since 
$R'(t) = Q \exp( iQ {\cal L} Q t) Q$, we formally introduce 
a complete set of eigenfunctions
of the hermitian (with respect to the Kubo scalar product) 
operator
$Q {\cal L} Q$:
\begin{equation}
Q {\cal L} Q \phi_\lambda = \lambda \phi_\lambda \, ,
\end{equation}
where all eigenvalues are real. Thus we can write:
\begin{eqnarray}
( A | R'(t) | A) &= &  ( Q A | e^{i Q{\cal L} Q t} | Q A ) 
\nonumber \\ 
& = &  
\sum_\lambda ( Q A | \phi_\lambda) e^{i\lambda t} (\phi_\lambda | Q A) \, .
\end{eqnarray}
The Laplace transform yields: 
\begin{equation}
LT[ ( A | R'(t) | A)](\omega) = \sum_\lambda \frac{1}{\omega - \lambda} 
| ( Q A | \phi_\lambda) |^2 \, ,
\end{equation}
with complex frequencies in the lower half-plane.
Since all the poles are located on 
the real axis, the Laplace transform is analytic for 
$\omega = \Omega - i \epsilon, \epsilon > 0$.
Furthermore:
\begin{eqnarray}
& & Im LT[ ( A | R'(t) | A)](\omega) \nonumber \\
& & =   \sum_\lambda 
\frac{\epsilon}{(\Omega-\lambda)^2 + \epsilon^2} 
| ( Q A | \phi_\lambda) |^2 \geq 0 \, .
\end{eqnarray}
In particular, provided the limit $\epsilon \searrow 0$ 
exists, one obtains for real $\omega$: 
\begin{equation}
Im LT[ ( A | R'(t) | A)](\omega) = \sum_\lambda 
\pi \delta(\omega-\lambda)
| ( Q A | \phi_\lambda) |^2 \geq 0
\end{equation}

Consider now a collection of phase space variables $A_i, i=1,..,l$ of 
identical time inversion parity. Then the 
real symmetric 
matrix $Im LT[ (A_i | R'(t) | A_j )](\omega)$ is positive semi-definite:  
since for arbitrary real numbers $y_i, i=1,..,l$ the spectrum of the 
autocorrelation function of
$Y = \sum_{i=1}^l y_i A_i$ is non-negative, one finds:
\begin{equation}\label{Onsager}
\sum_{i,j=1}^n y_i y_j Im LT[ (A_i | R'(t) | A_j) ](\omega) \geq 0 \, ,
\end{equation} 
which implies the property. For frequencies $\omega \to 0$, one obtains 
Onsager's relations, viz. the matrix of the kinetic coefficients is symmetric
with non-negative eigenvalues. Hence, \gl{Onsager} can be interpreted as 
the proper generalisation of Onsager's relations to finite frequencies.

\section{Transverse phonons and the Andersen-Pecora Approach in the low 
temperature limit}

If one makes the (incorrect) supposition that the memory
kernels of the Andersen-Pecora approach can be modeled by Debye 
relaxation processes consistent with their high-temperature Markov
approximation and their time reversal symmetry, this 
yields:
\begin{eqnarray}\label{E1}
M(\omega) = i \frac{A^2}{\tau} \frac{1}{1+ i \omega \tau} \, ,
\end{eqnarray} 
 
\begin{equation}
\hat{\eta}_s(\omega) = i \hat{\eta}_s^0 \frac{\tau}{1+ i \omega \tau} 
\, , \qquad \hat{\eta}_s^0 > 0 \, ,
\end{equation}
\begin{equation}\label{E3}
\lambda(\omega) = - \left[ i f_1 + f_2 \frac{\omega \tau}{1 + i \omega \tau}
\right] \, , \qquad f_1 , f_2 > 0 \, .
\end{equation}
The special form proposed for Eq. (\ref{E3}) derives from the 
fact that $\lambda(t)$ is an odd function of time. If we suppose it to be the
time derivative of $f(t)$, the auto-correlation function of some variable, 
$f_1$ 
is its $t=0$ value and we have chosen for its late-time evolution 
a smooth Debye-like behaviour. 

Let us admit that, in the $\omega \tau \gg 1$ regime, the 
value of $A$ is smaller than $1/\sqrt{2}$. Equation (\ref{E1}) then yields
a VH backscattering spectrum with a 
pseudo-Lorentzian line shape and a line width 
approximately equal to $\tau^{-1} A^2 (1- 2 A^2)^{-1/2}$. For $\omega$ 
larger than this line width, one can write
for the $q$-dependent part of Eq. (\ref{Andersen17}): 
\begin{eqnarray}\label{E4}
I(\vek{q},\omega) &= & 
-q^2 \left(\frac{\Lambda'}{m n \omega_0^2} \right)^2 
\frac{m^2 v^2}{\omega} 
\left[ Re \left( \frac{\omega \lambda(\omega)}{\omega- M(\omega)} \right)^2
\right] \nonumber \\
& & 
 \times Im \frac{1}{\omega^2 -q^2 \omega \eta_T(\omega)/mn } \, .
\end{eqnarray}
In this $\omega \tau \gg 1$ limit, this reads:
\begin{eqnarray}\label{E5}
I(\vek{q},\omega) &= & 
-q^2 \left(\frac{\Lambda'}{m n \omega_0^2} \right)^2 
\frac{m^2 v^2}{\omega}  \left( f_1 -f_2 \right)^2
 \nonumber \\
& & 
 \times Im \frac{1}{\omega^2 -q^2 c_T^2 + i\epsilon } \, ,
\end{eqnarray}
$\epsilon$ being a small positive quantity. Equation 
(\ref{E5}) does 
represent the Brillouin spectrum of a transverse phonon, but 
the square of its the velocity is given by:
\begin{eqnarray}
c_T^2 = \frac{\hat{\eta}_s^0}{mn} + \frac{\Lambda'}{mn \omega_0^2} 
(f_1-f_2)^2 \, .
\end{eqnarray}
In the spirit of the Andersen-Pecora approach, $\hat{\eta}_s(t)$ represents
the contribution of the molecular center-of-mass motion to the 
shear viscosity (see, nevertheless, the remark below Eq. (\ref{hateta})). 
Within the same spirit, $[\hat{\eta}_s^0/mn]^{1/2}$ should represent 
the contribution of the same motion to the transverse sound velocity. 
As announced, the bare  sound velocity, $[\hat{\eta}_s^0/mn]^{1/2}$, 
is renormalized by a term in $\Lambda'/\omega_0^2$, the 
signature of the rotation-translation coupling, but this
renormalisation  
leads to an unphysical increase of the sound velocity, instead of the expected
physical decrease.


\end{document}